\documentclass[journal]{IEEEtran}

\ifCLASSINFOpdf
\else
\fi

\usepackage{graphicx}
\usepackage{float}
\usepackage{amsmath}
\usepackage{stix}
\usepackage{mleftright,mathtools}
\DeclareMathAlphabet\mathbfcal{LS2}{stixcal}{b}{n}
\usepackage{mathrsfs}
\usepackage[mathscr]{euscript}
\usepackage{xspace}
\usepackage[super]{nth}
\usepackage[noadjust]{cite}
\usepackage[shortlabels, inline]{enumitem}
\usepackage{optidef}
\usepackage{amsthm}
\usepackage{amsfonts}
\usepackage{caption}
\usepackage{subcaption}
\theoremstyle{definition}

\usepackage{multirow}
\usepackage{bbm}

\usepackage{tabularx}
\usepackage{dirtytalk}
\usepackage{pgfplots}
\usepackage{pgfplotstable}
\usetikzlibrary{pgfplots.groupplots}
\usepackage[ruled,vlined]{algorithm2e}
\newcolumntype{C}[1]{>{\centering\arraybackslash}p{#1}}

\newcommand{\ie}{i.e.,\xspace}

\usepackage{color, xcolor, colortbl}

\newif\ifplotfigure
\plotfigurefalse 
\makeatletter
\def\subsubsection{\@startsection{subsubsection}{3}{\z@}%
                                   {1.5ex plus 1.5ex minus 0.5ex}%
                                   {0.5ex plus .ex minus 0ex}%
                                   {\normalfont\normalsize\it}}
\makeatother

\begin{document}

%
\title{DRL-Based RAN Slicing with Efficient Inter-Slice Isolation in Tactical Wireless Networks}
%
%
%

\author{\IEEEauthorblockN{Abderrahime Filali, \IEEEmembership{Member,~IEEE,}
Diala Naboulsi, \IEEEmembership{Senior Member,~IEEE,} and
Georges Kaddoum, \IEEEmembership{Senior Member, IEEE}}
}

\maketitle

\begin{abstract}
The next generation of tactical networks (TNs) is poised to further leverage the key enablers of 5G and beyond 5G (B5G) technology, such as radio access network (RAN) slicing and the open RAN (O-RAN) paradigm, to unlock multiple architectural options and opportunities for a wide range of innovative applications. RAN slicing and the O-RAN paradigm are considered game changers in TNs, where the former makes it possible to tailor user services to users’ requirements, and the latter brings openness and intelligence to the management of the RAN. In TNs, bandwidth scarcity requires a dynamic bandwidth slicing strategy. Although this type of strategy ensures efficient bandwidth utilization, it compromises RAN slicing isolation in terms of quality of service (QoS) performance. To deal with this challenge, we propose a deep reinforcement learning (DRL)-based RAN slicing mechanism that achieves a trade-off between efficient RAN bandwidth sharing and appropriate inter- and intra-slice isolation. The proposed mechanism performs bandwidth allocation in two stages. In the first stage, the bandwidth is allocated to the RAN slices. In the second stage, each slice partitions its bandwidth among its associated users. In both stages, the slicing operation is constrained by several considerations related to improving the QoS of slices and users that in turn foster inter- and intra-slice isolation. The proposed RAN slicing mechanism is based on DRL algorithms to perform the bandwidth sharing operation in each stage. We propose to deploy the mechanism in an O-RAN architecture and describe the O-RAN functional blocks and the main DRL model lifecycle management phases involved. We also develop three different implementations of the proposed mechanism, each based on a different DRL algorithm, and evaluate their performance against multiple baselines across various parameters.
\end{abstract}


%
\IEEEpeerreviewmaketitle

\vspace{-3mm}
\section{Introduction}
\IEEEPARstart{T}{he} next generation of tactical networks (TNs) is expected to have advanced capabilities to support a plethora of applications related to command, control, communication, intelligence, surveillance, and reconnaissance that will enhance the operation of the command-and-control structure. TNs have seized the opportunity to embrace 5G and beyond 5G (B5G) commercial technology to achieve this objective. Indeed, 5G and B5G networks are paving the way towards a robust, resilient, and secure communication infrastructure for TNs \cite{baeza2024new}. Embracing 5G and B5G technology enables TNs to leverage their key enablers, such as radio access network (RAN) slicing and the open RAN (O-RAN) paradigm. 

RAN slicing technology makes it possible to virtually divide a shared physical RAN into several logical networks called RAN slices \cite{elayoubi20195g}. Each RAN slice can have its own network performance indicators in accordance with the quality of service (QoS) requirements that apply to the service it provides. In TNs, RAN slicing makes it possible to handle a variety of tactical applications by devoting to each RAN slice the RAN resources it requires. Therefore, the integration of RAN slicing in TNs marks a radical shift towards flexible, scalable, and resource-optimized networks.

\begin{figure}[t]
	\includegraphics[width=0.8\linewidth]{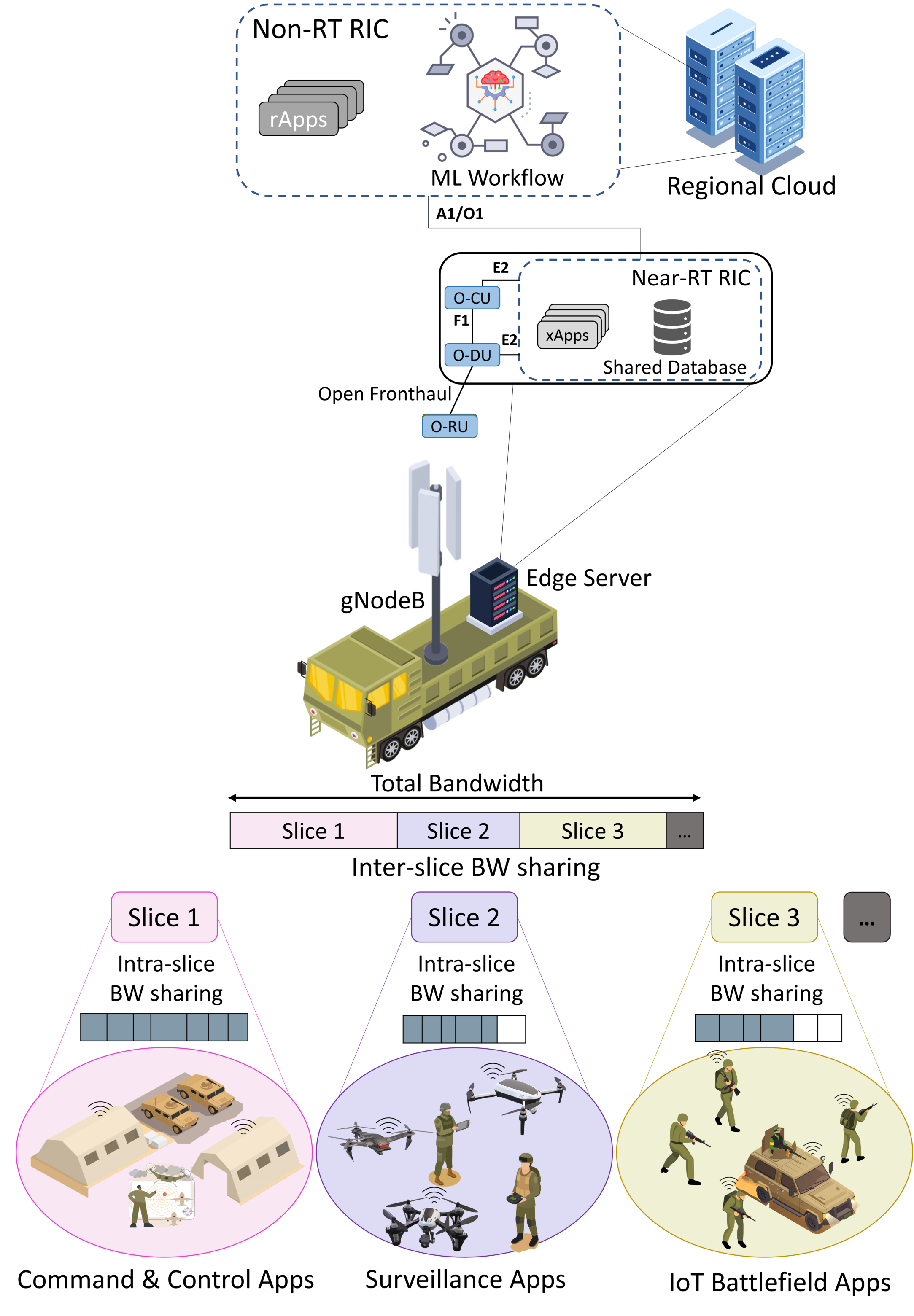}
	\caption{O-RAN system model for a TN environment.}
	\label{fig:01}
\end{figure}

To ensure the sustainable development of their RAN communication systems, TNs must become more receptive to existing RAN software and hardware products that have been developed by commercial technology suppliers. The flexibility to adopt third-party RAN solutions can be achieved through the O-RAN paradigm \cite{ORAN}. The main idea behind the O-RAN paradigm is to support RAN component intercompatibility and interoperability by making RAN hardware and software elements not locked to a specific supplier. The \mbox{O-RAN} paradigm enables TNs to take advantage of innovations developed by a wide range of suppliers to adapt their RAN infrastructure to cutting-edge technologies \cite{marques20235g}. Another game-changing capability empowered by the \mbox{O-RAN} paradigm is the automation of RAN operations management, such as RAN slicing, using machine learning (ML) \cite{azimi2022applications}. The desire for a more programmable RAN is driven by a pivotal component of the O-RAN architecture called the RAN intelligent controller (RIC) \cite{ORANWG2-Non-RT-RIC,ORANWG3-Near-RT-RIC}, which is shown in \mbox{Fig. \ref{fig:01}}. TNs can use the RIC to gather and analyze large volumes of data from the RAN. These data are leveraged to train and refine ML models embedded in the RIC, enabling the optimization of network operations and enhancing decision-making processes for command units.

\setlength{\textfloatsep}{0.1cm}
\setlength{\floatsep}{0.1cm}

RAN slicing and the O-RAN paradigm hold significant promises for enabling the next generation of TNs to deliver innovative services. However, implementing RAN slicing in TNs is complicated by several inherent challenges. First, TNs operate in a highly dynamic RAN environment where wireless channel transmission conditions are subject to constant and unpredictable changes. In addition, the scarcity of radio resources, such as bandwidth, represents a significant limitation for these networks. Adapting to this fast-changing and resource-constrained environment requires dynamic bandwidth sharing. However, this approach introduces challenges in maintaining slice isolation. In this study, we define RAN slicing isolation from the perspective of Quality of Service (QoS) performance. Specifically, isolating a slice refers to ensuring it is allocated the resources needed to meet its QoS requirements. This reduces the impact of fluctuations in the resource requirements of one slice on the QoS performance of the other slices. These issues highlight the significant challenges involved in implementing effective and reliable RAN slicing in TNs.


RAN slicing in TNs has been investigated in the literature \cite{yang2023game,castanares2022tactical,castanares2021slice}, but with many limitations, mainly the ignoring of isolation, user mobility and user QoS diversity in the RAN slicing schemes proposed. In non-TNs, RAN slicing isolation has been explored at different levels of granularity, namely the slice level and the user level. In addition, several resource- and performance-based metrics have been used to define isolation in a RAN \cite{dai2024ran,ghafouri2024multi,gholami2023mobile, abedin2022elastic, marabissi2019highly, zhang2022dynamic, tian2021wireless, azimi2021energy,azimi2024mobility, motalleb2022resource, yarkina2022multi, gijon2022data, bai2022latency,hou2023inter,hashemian2024analytical,akyildiz2024hierarchical}. Nevertheles, the schemes proposed focus mainly on satisfying isolation constraints, without regard to resource wastage. In other words, these schemes tolerate a slice or a user for being allocated more than the minimum amount of resources needed to meet their QoS requirements. This can be acceptable in non-TNs, where the bandwidth is more easily accessible than in TNs. In TNs, however, bandwidth is a very precious resource and should be meticulously allocated to avoid wastage. Moreover, wasting it can be detrimental to RAN slicing isolation. In fact, wasting the bandwidth in RAN slicing leads to frequent reconfiguration of the resources allocated to slices or users, which has an impact on isolation.

The novelty of this work is to propose a RAN slicing mechanism for O-RAN TNs. The objective of the proposed mechanism is to achieve a trade-off between efficient RAN bandwidth sharing and appropriate inter- and intra-slice isolation. The mechanism is called the bandwidth sharing and inter- and intra-slice isolation (BS-IISI) mechanism. The \mbox{BS-IISI} mechanism hierarchically performs bandwidth allocation in two stages, namely inter-slice bandwidth sharing and intra-slice bandwidth sharing. It ensures adequate isolation both at the inter- and intra-slice levels of granularity and in terms of resource and QoS performance. We define a utility function to calculate the degree of user satisfaction. It determines whether the bandwidth allocated to a user fulfills its QoS requirements and captures any resource wastage during bandwidth allocation. User satisfaction increases when a user’s achieved data rate gets closer to the required data rate threshold and decreases when it gets further from the required threshold (above or below). This makes it possible to allocate bandwidth efficiently and prevents resources from being wasted, which, in turn, contributes to isolation. We define slice resource reconfiguration constraints to enhance inter-slice isolation. These constraints ensure that resources can be removed from a slice that has excess resources and added to a slice that requires additional resources. This avoids unnecessary slice resource reconfiguration, which improves inter-slice isolation.

In both stages of RAN slicing, \ie inter- and intra-slice bandwidth sharing, the BS-IISI mechanism utilizes deep reinforcement learning (DRL) algorithms to perform bandwidth allocation. We propose to deploy the BS-IISI mechanism in an O-RAN architecture, which follows the specifications issued by the O-RAN Alliance Working Group (WG) 2. Integrating the BS-IISI mechanism in the O-RAN architecture allows for a high degree of automation and flexibility in RAN slicing operations to be able to instantly adapt to fluctuations in user traffic and the environmental conditions of TNs.

This paper's main contributions are summarized as follows:

\begin{itemize}
    \item We model the bandwidth allocation problem as a continuous optimization problem to maximize the user degree of satisfaction and ensure inter- and intra-slice isolation.

    \item We propose a DRL-based RAN slicing mechanism to solve the bandwidth allocation problem in each stage of RAN slicing.

    \item We propose to deploy the BS-IISI mechanism in an \mbox{O-RAN} architecture, and describe the O-RAN functional blocks involved and the key DRL model lifecycle management phases involved.

    \item We develop three different implementations of the BS-IISI mechanism, each based on a different DRL algorithm, and subsequently evaluate their performance against multiple baselines across various parameters.
\end{itemize}

The rest of the paper is structured as follows: Section II reviews recent related work, Section III outlines the system model, Section IV formulates the problem, Section V details the proposed RAN slicing mechanism, Section VI evaluates its performance, and Section VII concludes the paper.


\vspace{-3mm}
\section{Related Work}
In this section, we first review recent research pertaining to RAN slicing in TNs. Then, we present an overview of relevant works related to RAN slicing isolation and classify these works based on their level of isolation granularity level and the isolation metrics they consider.
\vspace{-4mm}
\subsection{RAN Slicing in Tactical Networks}
\vspace{-1mm}
In \cite{yang2023game}, the authors use game theory to model resource slicing in ad-hoc TNs as a non-cooperative game where slices compete for shared bandwidth to maximize throughput-based utility. They propose an iterative algorithm that converges to a Nash equilibrium, enabling slices to alternately select bandwidth to optimize their throughput. In \cite{castanares2022tactical}, the authors propose an algorithm to maximize device associations with tactical RAN slices using a priority mechanism based on the ratio of QoS to required bandwidth. In limited bandwidth scenarios, the algorithm reallocates resources from lower-priority slices to higher-priority ones to optimize device connectivity. In \cite{castanares2021slice}, the authors design a RAN slicing framework for managing the lifecycle of RAN slices in battlefield networks, with a slice controller that monitors QoS requirements and allocates necessary resources. These proposed approaches overlook inter-slice isolation, user mobility, and the clear definition of individual QoS requirements for slices.

\vspace{-2mm}
\subsection{RAN Slicing Isolation}
\vspace{-1mm}
Various works in the literature consider intra-cell isolation in their proposed RAN slicing schemes. These schemes can be classified by their isolation granularity and the isolation metrics they consider. Isolation granularity consists of two main levels, namely inter-slice isolation and intra-slice isolation. Several metrics have been used to gauge isolation, such as minimum and maximum thresholds related to allocated resources or QoS performance, as well as load fluctuation. These isolation metrics can be grouped into two classes: resource-based metrics and performance-based metrics. 

In \cite{dai2024ran}, the authors focus on ensuring inter-slice isolation by minimizing violations of service level agreements (SLAs). SLA violations are found to occur when the data rate or latency requirements of a slice's users are not met. To further enhance inter-slice isolation, resource sharing among slices is prohibited, even when a slice cannot fully use the resources allocated to it within a time window. In \cite{ghafouri2024multi}, a two-level scheme for RAN resource sharing is proposed to ensure slice isolation. At a high level, the system is found to guarantee the availability of RAN resources by monitoring the network and resource availability, and then assigning each user's request to the appropriate slice type. At the low level, based on the high-level decisions, the necessary resources were allocated to each user, which maintained isolation between slices. In \cite{gholami2023mobile}, RAN slice isolation is ensured by allocating to each slice a number of resource blocks (RBs) that is greater than the number requested and less than the maximum threshold value. In \cite{abedin2022elastic}, isolation is considered at the slice level and the user level subject to the constraints that a RB can be allocated to at most one slice at a time and each user must have a minimum number of RBs. In \cite{marabissi2019highly}, the objective is to maximize the number of RBs allocated to a slice, which is dependent on the slice’s traffic variation. This slice isolation metric is related to how much consideration is given to slice traffic variation in the RBs allocation process. In \cite{zhang2022dynamic,tian2021wireless}, isolation is defined as a ratio index of a user's QoS requirement to the QoS they actually achieve. This index is limited by a minimum value. Furthermore, each RAN slice’s throughput must not exceed a maximum threshold. 
In \cite{azimi2021energy}, RAN slice isolation is ensured by keeping the resources allocated to a given slice unchanged for a specific period of time. In other words, no RAN slice resource reconfiguration is performed during the time interval. In \cite{azimi2024mobility}, the authors adopted a similar approach to sustain RAN slice isolation at a higher level, involving the prediction of slice resource requirements on a large time scale to guarantee stable resource allocation. 
The work in \cite{motalleb2022resource} ensures RAN slice isolation by setting minimum and maximum thresholds for the number of RBs allocated to enhanced mobile broadband (eMBB) users and capping allocations for ultra-reliable low-latency communication (URLLC) and massive machine-type communication (mMTC) users. Additionally, user-level isolation is maintained by enforcing a minimum data rate threshold for each user.
In \cite{yarkina2022multi}, the infrastructure provider ensures the isolation of tenants' RAN slices in terms of data rate as long as the number of users associated with each slice does not exceed the contracted threshold. In \cite{gijon2022data}, the number of RBs a user requires is estimated from the throughput needed in the physical layer. Then, a slice receives the amount of resources that corresponds to the aggregation of the RBs estimated to schedule all its users. 

In \cite{hou2023inter}, the authors perform inter-slice resource allocation by reserving RBs for slices on long time scales based on predicted resource demands. To ensure a high level of resource isolation, the reservation process involves lending or borrowing resources between slices to maintain the necessary number of RBs. Next, a RB association operation is conducted to allocate the RBs to slices to uphold a high level of interference isolation. In \cite{hashemian2024analytical}, slice isolation is enhanced between interfering slices that operate on overlapping channels. This isolation is defined by the variation in user delay and throughput when traffic from interfering slices increases. An interference-aware channel allocation policy is proposed to reduce overlapping frequency channels. In \cite{akyildiz2024hierarchical}, eMBB and URLLC users are grouped based on their channel quality. Isolation is achieved by allocating RBs to each group in a way that maximizes the throughput for eMBB users while satisfying the delay requirements for URLLC users. \mbox{Table \ref{tab:1}} summarizes a classification of related work that investigated RAN slicing isolation.
\vspace{-3mm}
\begin{table}[h]
	\centering
    \captionsetup{aboveskip=0pt, belowskip=0pt}
	\caption{Classification of RAN slice isolation related work.}
	\label{tab:1}
    \begin{center}
	\begin{tabular}{|m{1.2cm}|m{1cm}|m{1cm}|m{1.4cm}|m{1.6cm}|}
		\hline
		\multirow{2}{*}{\textbf{Ref.}} & \multicolumn{2}{|c|}{\textbf{Isolation granularity}} & \multicolumn{2}{|c|}{\textbf{Isolation metric}}\\ 
         \cline{2-5}
         & \scriptsize{Inter-slice} & \scriptsize{Intra-slice} & \scriptsize{Resource-based} & \scriptsize{Performance-based} \\\hline
        \cite{gholami2023mobile,marabissi2019highly, akyildiz2024hierarchical} & \checkmark & & \checkmark & \\\hline
         \cite{abedin2022elastic,azimi2021energy} & \checkmark & \checkmark & \checkmark & \\\hline
         \cite{zhang2022dynamic} & \checkmark & \checkmark &  & \checkmark \\\hline
         \cite{tian2021wireless,azimi2024mobility,yarkina2022multi, hashemian2024analytical,ghafouri2024multi} & \checkmark & & & \checkmark \\\hline
         \cite{motalleb2022resource} & \checkmark & \checkmark & \checkmark & \checkmark \\\hline
        \cite{gijon2022data,bai2022latency} & & \checkmark & \checkmark & \\\hline
        \cite{hou2023inter,dai2024ran} & \checkmark & & \checkmark & \checkmark \\\hline
	\end{tabular}
 \end{center}
\end{table}
\vspace{-6mm}

Although RAN slicing isolation has been studied in previous works, there are still isolation-related research challenges that require more in-depth investigation. In this work, we propose a RAN slicing mechanism that ensures isolation both at the inter- and intra-slice levels of granularity and in terms of resource and QoS performance. Moreover, in previous research works, isolation has been achieved by allocating resources without any concern for their wastage. However, wasting resources can lead to poor isolation, since frequent resource reconfiguration is necessary when a large amount of resources are unused. The BS-IISI mechanism improves the efficiency of RAN slicing isolation while avoiding resource wastage.

\vspace{-3mm}
\section{System Model}
We consider a 5G O-RAN system composed of a single base station, \ie a 5G next-generation node B (gNodeB), \mbox{(see Fig. \ref{fig:01}}). The set of users associated with the gNodeB is denoted by $\mathscr{U} = \{1, \ldots, U\}$, where $|\mathscr{U}|=U$ 
.The gNodeB can support the deployment of a set of RAN slices denoted by $\mathscr{S} = \{1, \ldots, S\}$, where $|\mathscr{S}|=S$. Each RAN slice $s \in \mathscr{S}$ is designed to serve a subset of users denoted by $\mathscr{U}_s = \{1, \ldots, U_s\}$. Note that all users that are served by the same slice have the same QoS requirements. We assume that a user can be served by only one RAN slice, such that $\mathscr{U}_s \cap \mathscr{U}_{s^{'}}=\emptyset, \forall s \neq s{'}$ and $\cup_{ s \in \mathscr{S}} \mathscr{U}_s=\mathscr{U}$. We consider that the time dimension is divided into several time slots indexed \mbox{by $t$}. We assume that RAN slicing is performed at the beginning of each time slot $t$.

We consider that users move within the gNodeB's coverage area following the random waypoint (RWP) mobility model \cite{hyytia2007random}. Each user $u \in \mathscr{U}$ selects a random destination and moves to it at a constant speed $v_{u}$. Upon arrival, the user pauses for a random period $\tau_{u}$ before moving on. The user's velocity and pause time are assumed to follow the uniform distribution $\mathcal{U}(v_{min},v_{max})$ and $\mathcal{U}(0,\tau_{max})$, respectively. The variables $v_{min}$, $v_{max}$, and $\tau_{max}$ denote the user's minimum velocity, maximum velocity, and maximum pause time, respectively.

\vspace{-4mm}
\subsection{O-RAN architecture}
\vspace{-2mm}
Fig. \ref{fig:01} presents the considered O-RAN architecture that includes the following main components: the near-real-time (near-RT) RIC, the non-real-time (non-RT) RIC, the O-RAN radio unit (O-RU), the O-RAN distributed unit (\mbox{O-DU}), and the O-RAN central unit (O-CU). The near-RT RIC manages near-real-time control and optimization of O-RAN components and resources. By contrast, the non-RT RIC performs non-real-time management and optimization operations, handling tasks with latencies exceeding 1 second. In their turn, the \mbox{O-RU}, O-DU, and O-CU handle the gNodeB functions, with each unit responsible for specific tasks depending on the functional split option chosen, as defined by 3GPP \cite{3GPP_2}. The O-RAN Alliance has selected the 7.2x split option, with the O-RU handling lower-layer physical functions and the O-DU managing higher-layer tasks. This split ensures flexibility, with the precoding function located either in the O-DU \mbox{(Category A)} or in the \mbox{O-RU} (Category B) \cite{ORANWG4}. In this study, we selected Category A to keep the O-RU as simple as possible, as a less complex O-RU reduces size, weight, and power consumption, which are key factors in TNs. This architecture is ideal for TNs because it minimizes fronthaul bandwidth requirements, thus enabling low-latency and high-reliability communication in a resource-constrained environment.

To implement the O-RAN components in a network, the O-RAN Alliance defined six deployment scenarios \cite{alliance2020ran}. The choice of a deployment scenario depends on factors such as network resource constraints and the services the RAN is designed to deliver. In the present study, we selected \say{Scenario A} where (i) the O-RU is located at the cell site; (ii) the \mbox{O-DU}, O-CU, and near-RT RIC are hosted on an edge server; and (iii) the non-RT RIC is located in a regional cloud. The centralized control of the O-CU and O-DU enables efficient resource management, such as bandwidth allocation, which is critical for tactical operations requiring coordinated and real-time decision-making. Scenario A also reduces the need for extensive on-site equipment, aligning with the mobility and portability requirements of TNs. Furthermore, it supports scalability by allowing multiple O-RUs to connect to a centralized O-DU, which facilitates expanding the network as tactical situations evolve. This flexibility makes Scenario A ideal for large-scale deployments, such as covering multiple operational areas or supporting multiple tactical teams.

\vspace{-3mm}
\subsection{RAN Slicing Model}
\vspace{-0.9mm}
The gNodeB's total radio bandwidth, which is denoted \mbox{by $\mathscr{W}$}, is shared among the gNodeB's users. The slicing of $\mathscr{W}$ is performed hierarchically in two stages, namely inter-slice bandwidth sharing and intra-slice bandwidth sharing.  
\vspace{-1mm}
\subsubsection{Inter-Slice Bandwidth Sharing}~
\vspace{-4mm}

In this slicing stage, a resource scheduler assigns a portion $f_{s}^{t} \in [0,1]$ of the total bandwidth $\mathscr{W}$ to each slice $ s \in \mathscr{S}$ at time slot $t$. The bandwidth allocated to slice $s$ at time $t$ is denoted by $\mathscr{W}_{s}^{t}$ and calculated as follows:
\vspace{-2mm}
\begin{equation}\label{eq:sliceFraction}
    \mathscr{W}_{s}^{t} = f_{s}^{t}\mathscr{W}.
\end{equation}

To ensure that the gNodeB's total bandwidth is not exceeded and avoid the bandwidth being exploited by only a small subset of slices, we define the following two constraints:
\vspace{-2mm}
\begin{equation}\label{eq:slicesFractionsBudget}
    \sum_{ s=1}^{|\mathscr{S}|} f_{s}^{t} \leq 1
\end{equation}
and
\vspace{-3mm}
\begin{equation}\label{eq:sliceFractionMax}
    f^{min} \leq f_{s}^{t} \leq f^{max}, \forall s \in \mathscr{S}.
\end{equation}


\subsubsection{Intra-Slice Bandwidth Sharing}~
\vspace{-4mm}

In this slicing stage, each slice $s \in \mathscr{S}$ allocates to each of its users $u \in \mathscr{U}_s$ at time $t$ a portion $f_{u,s}^{t}$ of the \mbox{bandwidth $\mathscr{W}_{s}^{t}$} it was previously allocated in the inter-slice bandwidth sharing stage. We define the following two constraints for slice $s$:
\vspace{-2mm}
\begin{equation}\label{eq:usersFractionsBudget}
    \sum_{u=1}^{|\mathscr{U}_s|} f_{u,s}^{t} \leq 1
\end{equation}
and
\vspace{-3mm}
\begin{equation}\label{eq:userFractionMax}
    f_{s}^{min} \leq f_{u,s}^{t} \leq f_{s}^{max}, \forall u \in \mathscr{U}_s, \forall s \in \mathscr{S}.
\end{equation}

Constraint \eqref{eq:usersFractionsBudget} guarantees that the sum of the bandwidth portions allocated to the users served by slice $s$ does not exceed the slice's total bandwidth $\mathscr{W}_{s}^{t}$. Constraint \eqref{eq:userFractionMax} prevents the bandwidth $\mathscr{W}_{s}^{t}$ from being used by only a small subset of users.

\vspace{-3mm}
\subsection{Channel Model}
We assume that the gNodeB knows the channel state information of its associated users. The downlink channel gain between a user $u \in \mathscr{U}_s, \forall s \in \mathscr{S}$ and the gNodeB at time slot $t$ is denoted by $g_{u,s}^{t}$ and defined as follows:
\setlength{\abovedisplayskip}{0pt}
\begin{equation}\label{eq:gain}
    g_{u,s}^{t} = 10^{ - \frac{PL(d_{u,s})}{10}}|h_{u,s}^{t}|^{2},
\end{equation}
where $h_{u,s}^{t}$ is the small-scale fading coefficient, which is modeled as a complex Gaussian distribution. $PL(d_{u,s})$ is the path loss between user $u$ and the gNodeB. The path loss model considered in this work corresponds to the 3GPP specifications \cite{3GPP_1} and can be calculated as follows:
\setlength{\abovedisplayskip}{2pt}
\begin{equation}\label{eq:pathloss}
    PL(d_{u,s}) = 28 + 22\log_{10}(d_{u,s}) + 20\log_{10}(f_c) + \sigma_{SF},  
\end{equation}
where $d_{u,s}$ is the distance between user $u$ and the gNodeB, $f_c$ is the central frequency of the 5G band, and $\sigma_{SF}$ is the shadow fading that follows a normal distribution.

The downlink data rate achieved by user $u$ at time $t$ is calculated as follows: 
\begin{equation}\label{eq:datarate}
    r_{u,s}^{t} = f_{u,s}^{t}\mathscr{W}_{s}^{t}\log_{2}\left(1 + \frac{P g_{u,s}^{t}}{f_{u,s}^{t}\mathscr{W}_{s}^{t} N_{0}}\right),
\end{equation}
where $P$ is the downlink transmission power of the gNodeB and $N_{0}$ is the power of the additive white Gaussian noise (AWGN). We assume that $P$ is the same for all users. 

\vspace{-4mm}
\subsection{Degree of Satisfaction}
In this paper, we consider a user satisfied if the allocated bandwidth is sufficient to achieve a data rate that meets its QoS requirements. Since bandwidth is a limited resource, its efficient allocation is crucial to satisfying users' diverse QoS requirements. Inefficient bandwidth allocation can result in data rates below or above the required threshold. A data rate below the required threshold diminishes the QoS provided by the RAN slices, directly affecting users' degree of satisfaction. While exceeding the required data rate does not affect the QoS, it implies a waste of resources that could be allocated more effectively to other users facing bandwidth shortages. 

To accurately gauge users' degree of satisfaction, we need to define a utility function that effectively evaluates how well the bandwidth allocated meets user’s QoS requirements while capturing any resource wastage during the allocation process. For this reason, we model the degree of satisfaction of user $u \in \mathscr{U}_s, \forall s \in \mathscr{S}$ at time $t$ using Eq. \eqref{eq:sd_user} \cite{zhao2020network}:
\setlength{\abovedisplayskip}{0pt}
\begin{equation}\label{eq:sd_user}
    \Gamma_{u,s}^{t} =\frac{1-e^{\left(-\dfrac{\gamma_{u,s}^{t}}{\rho\dfrac{r_{u,s}^{t}}{R_{s}^{req}}}\right)}}{\varphi},
\end{equation}
where
\vspace{-3mm}
\begin{equation}\label{eq:sd_1}
    \gamma_{u,s}^{t} = \frac{ \left(\rho\dfrac{r_{u,s}^{t}}{R_{s}^{req}}\right)^{\xi}}
            {1+\left(\rho\dfrac{r_{u,s}^{t}}{R_{s}^{req}}\right)^{\xi}},
\end{equation}
$\rho$ and $\xi$ adjust the elasticity of the function. 
$R_{s}^{req}$ denotes the minimum data rate required by each user served by slice $s \in \mathscr{S}$. Note that all users served by the same slice have the same minimum data rate. $\varphi$ is a normalisation parameter that ensures that \mbox{$\Gamma_{u,s}^{t} \in [0,1]$} and is given by Eq. \eqref{eq:sd_normalisation_factor}:
\setlength{\abovedisplayskip}{0pt}
\begin{equation}\label{eq:sd_normalisation_factor}
    \varphi = 1-e^{-\dfrac{1}{(\xi-1)^{\dfrac{1}{\xi}}+ (\xi-1)^{\dfrac{1-\xi}{\xi}}}}.
\end{equation} 

According to Eq. \eqref{eq:sd_user}, with an increase of users' satisfaction, their achieved data rate gets closer to the required data rate and decreases as their achieved data rate gets farther from the required data rate (above or below). This helps to limit wasting resources by allocating users more than they require.

The normalized degree of satisfaction of a slice $s \in \mathscr{S}$ at time $t$ is defined as follows:
\vspace{-1mm}
\begin{equation}\label{eq:SD_slice}
   \Gamma_{s}^{t} = \frac{1}{|\mathscr{U}_{s}|}\sum_{u=1}^{|\mathscr{U}_{s}|}\Gamma_{u,s}^{t}.
\end{equation}

The total normalized degree of satisfaction of the entire system at time $t$ is defined as follows:
\vspace{-1mm}
\begin{equation}\label{eq:SD_total}
   \Gamma^{t} = \frac{1}{|\mathscr{S}|}\sum_{s=1}^{|\mathscr{S}|}\Gamma_{s}^{t}.
\end{equation}

\vspace{-5mm}
\subsection{Inter-Slice Isolation}
To deal with the fast-changing and resource-constrained tactical RAN environment, the resources allocated to RAN slices during the inter-slice bandwidth sharing stage need to be dynamically reconfigured. Indeed, some slices may require additional resources to maintain their users' degree of satisfaction at a high level. However, adjusting the bandwidth portions allocated to the RAN slices can adversely impact inter-slice isolation. Adding resources to or removing resources from a slice can reduce the degree of satisfaction of the slice's associated users, which affects isolation among slices.

We define the reconfiguration cost of a slice $s \in \mathscr{S}$ at \mbox{time $t$}, which is denoted by $\mathcal{C}_{s}^{t}$, as follows:
\vspace{-1mm}
\begin{equation}\label{eq:cost_slice}
     \mathcal{C}_{s}^{t}=
    \begin{cases}
        \Gamma_{s}^{t-1} - \Gamma_{s}^{t} & \text{if } \mathscr{W}_{s}^{t-1} \neq \mathscr{W}_{s}^{t} \text{ and } \Gamma_{s}^{t} < \Gamma_{s}^{t-1}\\
        0 & \text{ otherwise.}
    \end{cases}
\end{equation}

In Eq. \eqref{eq:cost_slice}, the reconfiguration cost of slice $s$ is considered if the slice underwent bandwidth reconfiguration at time $t$, \mbox{\ie $\mathscr{W}_{s}^{t-1} \neq \mathscr{W}_{s}^{t}$}, and the slice's degree of satisfaction is lower at time $t$ than it was at time $t-1$, \ie $\Gamma_{s}^{t} < \Gamma_{s}^{t-1}$. The value of the reconfiguration cost, $ \Gamma_{s}^{t-1} - \Gamma_{s}^{t}$, is proportional to the severity of the QoS deterioration. In other words, the more the slice's degree of satisfaction has deteriorated, the higher the reconfiguration cost, and vice versa.
The total normalized reconfiguration cost at time $t$ is defined as follows:
\vspace{-1mm}
\begin{equation}\label{eq:cost_total}
     \mathcal{C}^{t}=\frac{1}{|\mathscr{S}|}\sum_{s=1}^{|\mathscr{S}|}\mathcal{C}_{s}^{t}.
\end{equation}

In this study, we consider that isolating a slice refers to ensuring it is allocated the bandwidth it requires to meet its QoS performance. This implies reducing the impact of fluctuations in the resource demands of one slice on the QoS performance of the other slices. In other words, increasing or decreasing the portion of bandwidth allocated to a slice should not impact the other slices’ degree of satisfaction. To this end, we define slice resource reconfiguration constraints to efficiently reconfigure the bandwidth allocated to the slices
 and, in turn, ensure inter-slice isolation.

We consider slice $s$ to require additional resources at \mbox{time $t$} if one the following constraints is satisfied:

\textbf{\textit{Constraint 1:}}
\begin{itemize}
    \item almost all its allocated bandwidth $\mathscr{W}_{s}^{t-1}$ was used at \mbox{time $t-1$}, \ie $(1-\sum_{u=1}^{|\mathscr{U}_{s}|}f_{u,s}^{t-1}) \leq f^{min}$, and
    \item there is at least one unsatisfied user at time $t-1$, \mbox{\ie $\mathscr{U}_{s,uns}^{t-1} \neq \emptyset$}, where $\mathscr{U}_{s,uns}^{t-1} = \{ u \in \mathscr{U}_{s}, r_{u,s}^{t-1} \leq R_{s}^{req}\}$.
\end{itemize}

\textbf{\textit{Constraint 2:}}
\begin{itemize}
    \item its allocated bandwidth $\mathscr{W}_{s}^{t-1}$ was not fully used at \mbox{time $t-1$}, \ie $(1-\sum_{u=1}^{|\mathscr{U}_{s}|}f_{u,s}^{t-1}) \geq f^{min}$,
    \item $\mathscr{U}_{s,uns}^{t-1} \neq \emptyset$, and
    \item its remaining resources are not sufficient to meet the requirements of the unsatisfied users, \ie $f_{u^{*},s}^{t-1}|\mathscr{U}_{s,uns}^{t-1}| \geq (1-\sum_{u=1}^{|\mathscr{U}_{s}|}f_{u,s}^{t-1})$, where $f_{u^{*},s}^{t-1}$ is the portion of bandwidth allocated to the user $u^{*}$ with the lowest channel gain. We consider the amount of resources allocated to the user with the least channel gain in order to calculate the resources required in the worst-case scenario.
\end{itemize}

\textbf{\textit{Constraint 1}} indicates that although slice $s$ used almost all its allocated bandwidth $\mathscr{W}_{s}^{t-1}$ at time $t-1$, it did not satisfy all its users. \textbf{\textit{Constraint 2}} states that even if slice $s$ partitioned its bandwidth poorly among its users, its unused resources are not sufficient to accommodate all its users.

{If either \textbf{\textit{Constraint 1}} or \textbf{\textit{Constraint 2}} is satisfied for a slice $s \in \mathscr{S}$, we consider that this slice requires additional bandwidth at time $t$ and define this resource demand using the following binary parameter (see Eq. \eqref{eq:constr_y1}):
\vspace{-1mm}
\begin{equation}\label{eq:constr_y1}
    \vartheta_{s,1}^{t}=
    \begin{cases}
        1 & \text{if } s \text{ requires additional resources}\\
        0 & \text{otherwise}.
    \end{cases}
\end{equation}

We model whether a slice $s \in \mathscr{S}$ has experienced an increase in its allocated bandwidth portion $f_{s}^{t}$ at time $t$ using the following binary parameter (see Eq. \eqref{eq:constr_x1}):
\setlength{\abovedisplayskip}{4pt}
\begin{equation}\label{eq:constr_x1}
      \varkappa_{s,1}^{t}=
    \begin{cases}
        1 & \text{if } f_{s}^{t} \geq f_{s}^{t-1} \\
        0 & \text{otherwise}
    \end{cases}
\end{equation}

The relationship between $ \vartheta_{s,1}^{t}$ and $\varkappa_{s,1}^{t}$ can be expressed as a logical implication, \mbox{\ie $ \vartheta_{s,1}^{t} \implies \varkappa_{s,1}^{t} \equiv \bar{\vartheta}_{s,1}^{t} \vee \varkappa_{s,1}^{t}$}.}

We consider slice $s$ to have available resources at time $t$ if one of the following constraints is satisfied:

\textbf{\textit{Constraint 3:}}
\begin{itemize}
    \item all users associated with slice $s$ are satisfied, \mbox{\ie $\mathscr{U}_{s,uns}^{t-1}=\emptyset$, and}
    \item $\Gamma_{s}^{t-1} < \Gamma_{th}$, where $\Gamma_{th}$ is a slice satisfaction threshold.  
\end{itemize}

\textbf{\textit{Constraint 4:}}
\begin{itemize}
    \item $\mathscr{U}_{s,uns}^{t-1}=\emptyset$,
    \item $\Gamma_{s}^{t-1} \geq \Gamma_{th}$, and  
    \item there are available resources $(1-\sum_{u=1}^{|\mathscr{U}_{s}|}f_{u,s}^{t-1}) \geq f_{s}^{min}$.
\end{itemize}

\textbf{\textit{Constraint 3}} indicates that all users associated with \mbox{slice $s$} have achieved the minimum data rate required, but their average degree of satisfaction is low. This means that the bandwidth $\mathscr{W}_{s}^{t-1}$ allocated to slice $s$ at time $t-1$ has been wasted during intra-slice bandwidth allocation. In other words, some users associated with slice s are allocated more bandwidth than they require. \textbf{\textit{Constraint 4}} ensures that there are unused resources in the slice.

If either \textbf{\textit{Constraint 3}} or \textbf{\textit{Constraint 4}} is satisfied for a slice $s \in \mathcal{S}$, we consider that this slice to have unused bandwidth at time $t$ and define this resource excess using the following binary parameter (see Eq. \eqref{eq:constr_y2}). 
\begin{equation}\label{eq:constr_y2}
    \vartheta_{s,2}^{t}=
    \begin{cases}
        1 & \text{if } s \text{ has available resources}\\
        0 & \text{otherwise}.
    \end{cases}
\end{equation}
We model whether a slice $s \in \mathcal{S}$ has experienced a decrease in its allocated bandwidth portion $f_{s}^{t}$ at time $t$ using the binary parameter shown in Eq. \eqref{eq:constr_x2}: 
\begin{equation}\label{eq:constr_x2}
      \varkappa_{s,2}^{t}=
    \begin{cases}
        1 & \text{if } f_{s}^{t} \leq f_{s}^{t-1} \\
        0 & \text{otherwise}
    \end{cases}
\end{equation}
The relationship between $ \vartheta_{s,2}^{t}$ and $\varkappa_{s,2}^{t}$ can be defined as a logical implication, \ie $ \vartheta_{s,2}^{t} \implies \varkappa_{s,2}^{t} \equiv  \bar{\vartheta}_{s,2}^{t} \vee \varkappa_{s,2}^{t}$.

\vspace{-3mm}
\section{Problem Formulation}
\vspace{-1mm}
In order to satisfy the QoS of users, we need to achieve a trade-off between efficient RAN bandwidth sharing among slices and appropriate inter-slice isolation. Therefore, we should maximize each slice's degree of satisfaction while minimizing its reconfiguration cost. We transform the two objectives into a weighted sum by introducing a weight factor $\alpha \in [0,1]$. The global bandwidth allocation optimization problem is formulated as follows: 
\begin{maxi!}[2]
	{f_{s}^{t},f_{u,s}^{t}}{\alpha \, \Gamma^{t} - (1-\alpha) \, \mathcal{C}^{t} \label{eq:obj1}}
	{\label{eq:opt1}}{}
    \addConstraint{\sum_{s=1}^{|\mathscr{S}|}f_{s}^{t}}{\leq 1 \label{eq:obj1_Cb}} 
    \addConstraint{f^{min} \leq f_{s}^{t}}{\leq f^{max}, \forall s \in \mathscr{S} \label{eq:obj1_Cc}}
	\addConstraint{\sum_{u=1}^{|\mathscr{U}_{s}|}f_{u,s}^{t}}{\leq 1, \forall s \in \mathscr{S} \label{eq:obj1_Cd}}
	\addConstraint{f_{s}^{min} \leq f_{u,s}^{t}}{\leq f_{s}^{max}, \forall u \in \mathscr{U}_s, \forall s \in \mathscr{S} \label{eq:obj1_Ce}}
    \addConstraint{ r_{u,s}^{t}}{\geq R_{s}^{req}, \forall u \in \mathscr{U}_s, \forall s \in \mathscr{S} \label{eq:obj1_Cf}}
    \addConstraint{ \bar{\vartheta}_{s,1}^{t} \vee \varkappa_{s,1}^{t}}{ = 1, \forall s \in \mathscr{S} \label{eq:obj1_Cg}}
    \addConstraint{ \bar{\vartheta}_{s,2}^{t} \vee \varkappa_{s,2}^{t}}{ = 1, \forall s \in \mathscr{S} \label{eq:obj1_Ch}}
    \addConstraint{ \varkappa_{s,1}^{t} \oplus \varkappa_{s,2}^{t}}{ = 1 , \forall s \in \mathscr{S} \label{eq:obj1_Ci}}
	\addConstraint{f_{s}^{t}, f_{u,s}^{t} \in [0,1], \forall u \in \mathscr{U}_s, \forall s \in \mathscr{S} \label{eq:obj1_Cj}}
    \addConstraint{\varkappa_{s,1}^{t}, \varkappa_{s,2}^{t}, \vartheta_{s,1}^{t}, \vartheta_{s,2}^{t}  \in \{0,1\}, \forall s \in \mathscr{S} \label{eq:obj1_Ck}}.
\end{maxi!}
Constraint \eqref{eq:obj1_Cb} ensures that the sum of the bandwidth portions allocated to the slices does not exceed the total bandwidth of the system. Constraint \eqref{eq:obj1_Cd} ensures that each slice does not receive more bandwidth that it has been allocated. Constraints \eqref{eq:obj1_Cc} and \eqref{eq:obj1_Ce} set minimum and maximum thresholds for the bandwidth portions allocated to slices and users, respectively. 
Constraint \eqref{eq:obj1_Cf} ensures that the data rate achieved by each user is above the minimum threshold. Constraints \eqref{eq:obj1_Cg} and \eqref{eq:obj1_Ch} ensure that the logical implication relationships between $\varkappa_{s,1}^{t}$ and $\vartheta_{s,1}^{t}$ and between $\varkappa_{s,2}^{t}$ and $\vartheta_{s,2}^{t}$ are maintained, respectively. Constraint \eqref{eq:obj1_Ch} implicitly includes $\Gamma_{th}$, making the solution space constrained by the degree of slice satisfaction. Constraint \eqref{eq:obj1_Ci} prevents the unfeasible slice resource allocation scenario in which a slice requires resources and at the same time has available resources. Constraint \eqref{eq:obj1_Cj} presents the optimization variables. 
Constraints \eqref{eq:obj1_Cc}, \eqref{eq:obj1_Cg}, and \eqref{eq:obj1_Ch} preserve inter-slice isolation. Constraint \eqref{eq:obj1_Cc} ensures each slice receives a minimum amount of resources and prevents resources from being monopolized by a subset of slices. Constraint \eqref{eq:obj1_Cg} ensures that resources that are added to a slice are allocated to a slice that needs them. Constraint \eqref{eq:obj1_Ch} ensures that resources that are removed from a slice are taken from a slice that has unused resources. As a result, unnecessary slice resource reconfiguration is avoided, which in turn promotes inter-slice isolation.  Constraints \eqref{eq:obj1_Ce} and \eqref{eq:obj1_Cf} preserve intra-slice isolation during dynamic resource sharing among users in a given slice. Isolation is guaranteed in terms of resources by Eq. \eqref{eq:obj1_Ce}, and performance by Eq. \eqref{eq:obj1_Cf}.

\vspace{-3mm}
\section{Bandwidth Sharing Inter-Slice and Intra-Slice Isolation Mechanism}
We leveraged DRL techniques to address the RAN slicing problem at each bandwidth allocation stage. DRL algorithms can be classified into model-based and model-free approaches. Model-based algorithms involve learning an explicit model of the environment, including the transition function, to guide action selection. In contrast, model-free algorithms rely solely on experience, learning to associate optimal actions with specific states without explicit knowledge of the environment.

The choice between model-based and model-free algorithms depends on the problem being investigated. Model-based algorithms are well-suited for fixed environments, such as factory robots or chess games, while model-free algorithms are better for scenarios with unknown or complex dynamics. In this work, we opt for model-free algorithms due to: (i) the highly dynamic nature of the RAN environment and its permanent variations, and (ii) the lack of a reliable environmental representation during RAN slicing. Each bandwidth allocation problem in each RAN slicing stage is formulated as a Markov decision process (MDP).

\vspace{-3mm}
\subsection{MDP Formulation of Inter-Slice Bandwidth Allocation}
\vspace{-1mm}
In this slicing stage, we consider that a global agent is responsible for allocating portions of the entire system's bandwidth to the slices. 
\vspace{-2mm}
\subsubsection{The State Space}~
\vspace{-4mm}

To choose an action at time $t$, the global agent observes how the entire system performed following its decision at time $t-1$. The global agent's observation at time $t$ is defined as follows:
\begin{equation}\label{eq:state_global}
    \mathscr{O}_{g}^{t} = \langle \boldsymbol{\Upsilon^{t-1}, \mathscr{V}_{1}^{t-1}, \mathscr{V}_{2}^{t-1}, \mathscr{F}^{t-1}} \rangle,
\end{equation}
where $\boldsymbol{\Upsilon^{t-1}} = (\Gamma_{s}^{t-1} : s \in \mathscr{S})$ represents the slices' degree of satisfaction at time $t-1$. $\boldsymbol{\mathscr{V}_{1}^{t-1}} = (\vartheta_{s,1}^{t-1} : s \in \mathscr{S})$ \mbox{and $\boldsymbol{\mathscr{V}_{2}^{t-1}} = (\vartheta_{s,2}^{t-1} : s \in \mathscr{S})$} denote the resource reconfiguration constraints at time $t-1$ related to adding resources to and removing resources from the slices, respectively. \mbox{$\boldsymbol{\mathscr{F}^{t-1}} = (f_{s}^{t-1} : s \in \mathscr{S})$ }represents the portions of bandwidth $\mathscr{W}$ that were allocated to the slices at time $t-1$.

\vspace{-2mm}
\subsubsection{The Action Space}~
\vspace{-4mm}

To perform slice bandwidth allocation, the global agent must decide what portion of the entire system's \mbox{bandwidth $\mathscr{W}$} to allocate to each slice. The global agent's action space is defined as follows:
\vspace{-2mm}
\begin{equation}\label{eq:action_global}
    \mathscr{A}_{g}^{t} = [f^{min},f^{max}]^{|\mathscr{S}|}.
\end{equation}

An action $a_{g}^{t} \in \mathscr{A}_{g}^{t}$ is a row vector, where \mbox{$a_{g}^{t} = [f_{1}^{t},f_{2}^{t},\ldots,f_{|\mathscr{S}|}^{t}]$}.

\setlength{\abovedisplayskip}{0pt}
\subsubsection{The Reward Function}~
\vspace{-4mm}

The main objective of inter-slice bandwidth allocation is to maximize the total degree of satisfaction of the entire system, but not at the expense of inter-slice isolation. Accordingly, the reward the global agent receives at time $t$ when it chooses an action $a_{g}^{t} \in \mathscr{A}_{g}^{t}$ is related to: (i) upholding the system's total bandwidth constraint defined in Eq. \eqref{eq:obj1_Cb} as well as the minimum and maximum slice bandwidth portion thresholds defined in Eq. \eqref{eq:obj1_Cc}, (ii) satisfying the slice reconfiguration constraints defined in Eqs. \eqref{eq:obj1_Cg} and \eqref{eq:obj1_Ch}, and (iii) $\Gamma^{t}$, and $\mathcal{C}^{t}$.
When the global agent selects an action that violates at least one of constraints \eqref{eq:obj1_Cb}, \eqref{eq:obj1_Cc}, \eqref{eq:obj1_Cg} and \eqref{eq:obj1_Ch}, the action is considered as an invalid action. The reward the global agent receives at time $t$ is defined as follows:
\vspace{1mm}
\begin{equation}\label{eq:reward_global}
    \mathscr{R}_{g}^{t} = 
         \begin{cases}
        \alpha \, \Gamma^{t} - (1-\alpha) \, \mathcal{C}^{t} & \text{if } a_{g}^{t} \text{ is valid}\\
        -1 & \text{if } a_{g}^{t} \text{ is invalid}.
    \end{cases}
\end{equation}
\vspace{-1mm}
When the global agent chooses a valid action, it receives the value of the objective function defined in Eq. \eqref{eq:obj1}. On the other hand, if it chooses an invalid action, it is penalized with a negative reward to discourage it from choosing the same action again in the future.

\vspace{-5mm}
\subsection{MDP Formulation of Intra-Slice Bandwidth Allocation}
In this RAN slicing stage, we consider that each slice is assigned an agent that allocates portions of the slice's bandwidth to the slice's associated users. 

\vspace{-2mm}
\subsubsection{The State Space}~
\vspace{-4mm}

The state of the environment that is observed by \mbox{slice $s$'s} agent at time $t$ is defined as follows:
\vspace{0mm}
\begin{equation}\label{eq:state_slice}
    \mathscr{O}_{s}^{t} = \langle \boldsymbol{\mathcal{G}_{s}^{t}} \rangle,
\end{equation}
\vspace{0mm}
where $\boldsymbol{\mathcal{G}_{s}^{t}}=(g_{u,s}^{t}: u \in \mathscr{U}_s)$ represents the channel gain between the gNodeB and the users associated with slice $s$ at time $t$. Agent $s$'s observation at time $t$ is given by the row vector $[g_{1,s}^{t},g_{2,s}^{t},\ldots, g_{|\mathscr{U}_{s}|,s}^{t}]$. At each time $t$, the positions of the users associated with slice $s$ vary. 
This leads to variation in the users' channel gain values in accordance with Eqs. \eqref{eq:gain} and \eqref{eq:pathloss}. Hence, agent $s$'s observation varies at each time $t$ depending on the dynamics of the wireless channel.

\vspace{-2mm}
\subsubsection{The Action Space}~
\vspace{-4mm}

To perform intra-slice bandwidth allocation, slice $s$'s agent must observe its environment and decide what portion of slice $s$'s allocated bandwidth $\mathscr{W}_{s}^{t}$ to assign to its users. Slice $s$ agent's action space at time $t$ is defined as follows:
\setlength{\abovedisplayskip}{0pt}
\begin{equation}\label{eq:action_slice}
    \mathscr{A}_{s}^{t} = [f_{s}^{min},f_{s}^{max}]^{|\mathscr{U}_s|}.
\end{equation}

An action $a_{s}^{t} \in \mathscr{A}_{s}^{t}$ is a row vector, where \mbox{$a_{s}^{t} = [f_{1,s}^{t},f_{1,s}^{t},\ldots,f_{|\mathscr{U}_s|,s}^{t}]$}.

\vspace{-2mm}
\subsubsection{The Reward Function}~
\vspace{-1mm}
The reward function represents the agent's objective of allocating sufficient resources to meet users' QoS requirements. The reward value depends on the degree of satisfaction of the associated users. Intra-slice bandwidth allocation is deemed successful if it adheres to constraints \eqref{eq:obj1_Cd} and \eqref{eq:obj1_Ce}. Actions violating these constraints are considered invalid. The reward for slice $s$'s agent at time $t$ is defined as:

\vspace{-2mm}
\begin{equation}\label{eq:reward_slice}
    \mathscr{R}_{s}^{t} = 
         \begin{cases}
        \Gamma_{s}^{t} & \text{if } a_{s}^{t} \text{ is valid}\\
        -1 & \text{if } a_{s}^{t} \text{ is invalid}.
    \end{cases}
\end{equation}
\vspace{-2mm}

\vspace{-4mm}
\subsection{On-Policy and Off-Policy DRL}
\setlength{\abovedisplayskip}{0pt}
Model-free DRL algorithms are categorized as on-policy or off-policy. During training, the agent uses a behavior policy $\pi_{B}$ to select actions based on observed environmental states and collect data. It then employs a target policy $\pi_{T}$ to update Q-values by evaluating actions based on rewards. In on-policy algorithms, the behavior and target policies are the same, whereas in off-policy algorithms, they differ.


There are advantages and drawbacks to each class of algorithms. For instance, off-policy algorithms are more likely to find a global optimum solution, while on-policy algorithms may get trapped in a local optimum. On the other hand, on-policy algorithms are less computationally complex than off-policy algorithms since $\pi_{B}$ and $\pi_{T}$ are the same. 

On-policy and off-policy algorithms have been widely exploited to solve RAN slicing problems \cite{filali2022dynamic,abouaomar2022federated}. Both classes of algorithms have proven effective in RAN resource allocation problems. Therefore, determining the most appropriate algorithm from the two classes depends on the use case. In this work, we have chosen algorithms from both classes \mbox{- the} off-policy algorithms TD3 \cite{fujimoto2018addressing} and DDPG \cite{lillicrap2015continuous}, and the on-policy algorithm PPO \cite{schulman2017proximal}, to evaluate how they perform with the proposed RAN slicing mechanism. We designed three implementations of the proposed BS-IISI mechanism, each of which is based on a different DRL algorithm. For instance, in the TD3-based BS-IISI implementation, the TD3 algorithm is used in both the inter-slice bandwidth sharing stage and the intra-slice bandwidth sharing stage.

DDPG and TD3 are actor-critic algorithms that use deterministic policies. In DDPG, the actor network chooses actions based on the current policy, while the critic network estimates a Q-value to evaluate the actions taken by the actor. TD3 follows the same strategy, but uses two critic networks to compute the minimum predicted value for target updates. Both DDPG and TD3 use experience replay and target networks for stability. By contrast, PPO is an actor-critic algorithm that uses stochastic policies, where the actor network produces a probability distribution over actions instead of deterministic outputs, and the critic network evaluates the expected return of the chosen policy. PPO optimizes a surrogate objective with a clipping mechanism to constrain policy updates, ensuring stable and reliable learning.

\vspace{-3.5mm}
\subsection{BS-IISI Deployement in an O-RAN \mbox{Architecture}}

The pseudo-code of the proposed BS-IISI mechanism is presented in Algorithm \ref{alg:BS-IISI}. Inter-slice bandwidth allocation precedes intra-slice bandwidth allocation, and both processes are performed at the beginning of each time slot $t$. Prior to theses processes, we verify whether any RAN slices need additional resources. This can be done by observing the parameter $ \vartheta_{s,1}^{t-1} \, \forall s \in \mathscr{S}$ in line 1. If $\vee_{s \in \mathscr{S}} \vartheta_{s,1}^{t-1}=1$, the global agent utilizes its trained DRL model to perform RAN slicing by allocating a portion of the system's bandwidth to each RAN slice, \mbox{(see line 2)}. When none of the RAN slices requires additional resources, each slice maintains the bandwidth portion it obtained at \mbox{time $t-1$}, such that \mbox{$f_{s}^{t-1}=f_{s}^{t}$}, \mbox{(see line 6)}. Verifying condition $\vee_{s \in \mathscr{S}} \vartheta_{s,1}^{t-1}=1$ makes it possible to efficiently perform inter-slice bandwidth allocation, since it is triggered only when there is a lack of resources in slices. This reduces the cost of reconfiguring slice resources, which in turn enhances inter-slice isolation.

Once the system's bandwidth has been allocated to the slices, it's time for intra-slice bandwidth allocation. Each \mbox{slice $s$'s} agent uses its DRL inference model to allocate to each of the slice's associated users a portion $f_{u,s}^{t}$ of the \mbox{bandwidth $\mathscr{W}_{s}^{t}$} it was previously allocated by the global agent, (see lines 4 and 8). Then, in lines 5 and 9, each slice agent publishes $\Gamma_{s}^{t}, \vartheta_{s,1}^{t}, \vartheta_{s,2}^{t}$ in the shared database. 
\begin{algorithm}[t]
\KwIn{$\mathscr{W},  \mathscr{S}, \mathscr{U}$}
\KwOut{$f_{s}^{t}, f_{u,s}^{t}, \forall u \in \mathscr{U}_s, \forall s \in \mathscr{S}$}
\ForEach{time $t$}{
    \nl \eIf{$\vee_{s \in \mathscr{S}} \vartheta_{s,1}^{t-1}=1$}{
     \nl The global agent uses its trained DRL model to choose $f_{s}^{t}, \forall s \in \mathscr{S}$\;
     \nl Each agent $s \in \mathscr{S}$: \\
      \nl  \quad uses its trained model to select $f_{u,s}^{t}, \forall u \in \mathscr{U}_s$\;
      \nl  \quad publishes $\Gamma_{s}^{t}, \vartheta_{s,1}^{t}, \vartheta_{s,2}^{t}$ in the shared database\;
    } {
    \nl $f_{s}^{t-1}=f_{s}^{t}, \forall s \in \mathscr{S}$ \;
    \nl Each agent $s \in \mathscr{S}$: \\
      \nl  \quad uses its trained model to select $f_{u,s}^{t}, \forall u \in \mathscr{U}_s$\;
      \nl  \quad publishes $\Gamma_{s}^{t}, \vartheta_{s,1}^{t}, \vartheta_{s,2}^{t}$ in the shared database\;
      }
}
\caption{BS-IISI Mechanism}
\label{alg:BS-IISI}
\end{algorithm}

\begin{figure}[t]
\captionsetup{aboveskip=0pt, belowskip=0pt}
	\centering \includegraphics[width=0.8\linewidth]{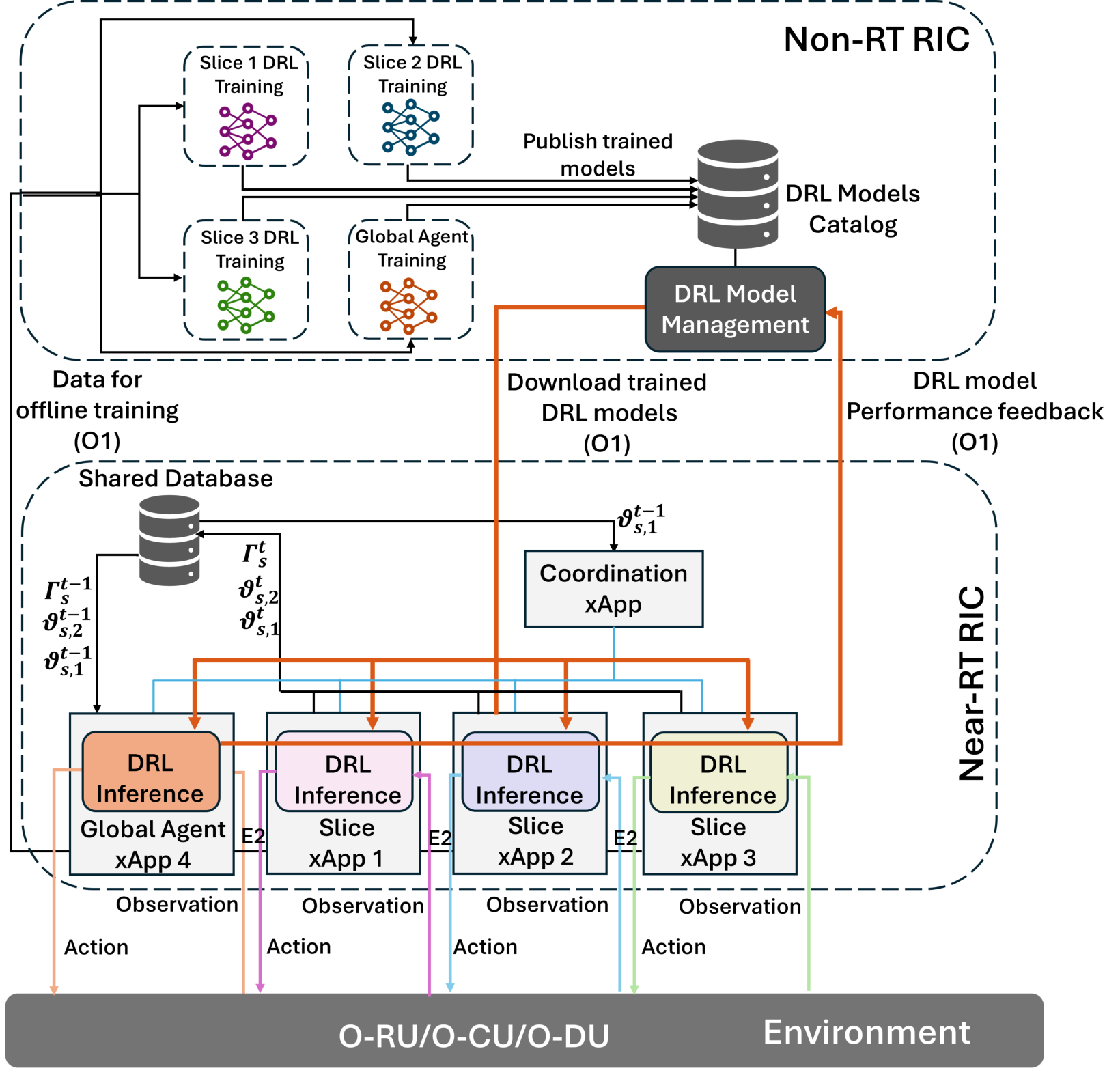}
	\caption{BS-IISI mechanism in a O-RAN framework.}
	\label{fig:02}
    \vspace{-1mm}
\end{figure}

Fig. \ref{fig:02} illustrates the use of the BS-IISI mechanism in a high-level O-RAN architecture. In \cite{ORANWG2}, the O-RAN Alliance WG2 outlined various deployment scenarios for RL-based mechanisms. We opted for \say{Scenario 1.4} with offline learning, as it is well aligned with the stringent requirements of TNs. In this scenario, the training host is located in the non-RT RIC, while the inference host resides in the near-RT RIC. This setup addresses key constraints, including latency and limited computing and storage resources frequently encountered in tactical environments. Specifically, both inter-slice and intra-slice bandwidth allocation should occur on a short time scale, requiring that inference models operate within the near-RT RIC. The non-RT RIC, with access to abundant computing and storage resources, handles training processes without overloading field-deployed infrastructure. This architecture ensures efficient resource utilization and low-latency operation, thus meeting the dynamic and mission-critical demands of TNs.


The inference models of the global agent and each RAN slice agent are each deployed in an xApp. Each agent receives observation data and executes bandwidth slicing actions through the E2 interface. Following intra-slice bandwidth allocation, each slice $s$'s agent publishes its satisfaction degree $\Gamma_{s}^{t}$, and parameters $\vartheta_{s,1}^{t}$ and $\vartheta_{s,2}^{t}$ to a shared database accessible by all xApps. The global agent uses information shared by slice agents for inter-slice bandwidth allocation. A coordinating xApp ensures that the condition $\vee_{s \in \mathscr{S}} \vartheta_{s,1}^{t-1}=1$ is satisfied before each allocation. At the beginning of each time slot $t$, the coordinating xApp: (i) reads the $\vartheta_{s,1}^{t-1}$ values from the shared database, (ii) checks $\vee_{s \in \mathscr{S}} \vartheta_{s,1}^{t-1}=1$, and (iii) decides whether the global agent proceeds with inter-slice bandwidth allocation. The Datasets collected during inference are sent to DRL training hosts for retraining through the O1 interface. Once a model is trained, tested, and validated, it is published in the model catalog. During inference, slice agents permanently report model performance to the model management entity, which decides whether an inference model needs to be updated. When an inference model suffers from a serious degradation in performance, the model management entity: (i) terminates the model, (ii) selects a new trained model from the catalog, and (iii) deploys it to the \mbox{near-RT RIC.}

\vspace{-4mm}
\section{Simulation Results}
\subsection{Simulation Settings}
\vspace{-1mm}
The proposed BS-IISI mechanism is designed to meet the QoS requirements of any tactical RAN slice. However, due to the lack of standardized QoS specifications for TN applications, we adopted a classification approach based on the 5G application categories. Accordingly, we consider a RAN configuration scenario in which the three main 5G services -- eMBB, URLLC, mMTC -- are deployed in a single gNodeB. These services are provided by three RAN slices, namely the eMBB slice, the URLLC slice and the mMTC slice. The gNobeB covers an area measuring $(500 m \times 500 m)$. Users
move according to the RWP model within the gNodeB's coverage area. Each slice has its own associated users. All the RAN simulation parameters are summarized in Table \ref{tab:2}.

We implement the proposed BS-IISI mechanism based on DRL algorithms. In fact, we developed three different implementations of the BS-IISI mechanism using three different DRL algorithms, namely TD3, DDPG, and PPO. For instance, in the TD3-based implementation, the TD3 algorithm is used in both bandwidth slicing stages, \ie the inter-slice bandwidth sharing stage and the intra-slice bandwidth sharing stage. All algorithms were implemented in Python using the Pytorch framework. For each algorithm, there were 1000 independent random realizations, using the inference models of the DRL algorithms considered, which were then averaged. Table \ref{tab:3} summarizes the hyperparameter values used in each DRL algorithm and bandwidth slicing stage. 

We compared the proposed BS-IISI mechanism with the following four baselines: TD3 without isolation constraints (TD3-WIC), DDPG-WIC, PPO-WIC , and the resource sharing and slice isolation based on an iterative process (RSSI-IP) \cite{yarkina2022multi}. 

In the TD3-WIC, DDPG-WIC, and PPO-WIC implementations, the BS-IISI mechanism operates without inter- and intra-slice isolation constraints defined in Eq. \eqref{eq:obj1_Cc}, \eqref{eq:obj1_Ce}, \eqref{eq:obj1_Cg}, and \eqref{eq:obj1_Ch}. The RSSI-IP mechanism addresses the RAN slicing problem by formulating it as a convex optimization problem to maximize user data rates. This mechanism uses an iterative process to achieve flexible performance isolation among slices while ensuring fair resource allocation. Slice performance isolation is maintained as long as the number of users does not exceed a contracted threshold between the slice tenant and the infrastructure provider. When some slices do not fully use their contracted resources, the excess resources are allocated to other slices whose demands exceed their thresholds. 

We opted for RSSI-IP as a benchmark since its objectives closely align with those defined in this work, notably in achieving efficient dynamic RAN resource sharing and robust slice isolation. In addition, in \cite{yarkina2022multi}, slice isolation is defined in accordance with our perspective, focusing on QoS performance, where isolating a slice refers to reducing the impact that fluctuations in the resource requirements of one slice have on the QoS performance of other slices. Furthermore, the RSSI-IP mechanism operates under assumptions that match our scenario, namely: i) a single cell RAN environment is considered; ii) each user is served by only one RAN slice; iii) each RAN slice supports a single service type with identical QoS requirements for its users; and iv) the data rate achieved by the slice's users should meet a minimum data rate threshold.

\begin{table}[t]
	\centering
    \captionsetup{aboveskip=0pt, belowskip=0pt}
	\caption{RAN parameters.}
	\label{tab:2}
    \begin{center}
	\begin{tabular}{|m{3.5cm}|m{4.2cm}|}
		\hline
		\textbf{Parameter} & \textbf{Value} \\ \hline
		
		\multicolumn{1}{|l|} { Total number of slices} & 3 \\\hline
        \multicolumn{1}{|l|} {Max. number of users per slice} & eMBB: 20; \newline URLLC:70;\newline mMTC:210. \\\hline
		\multicolumn{1}{|l|} {$\mathscr{W}, P, N_{0}, f_{c}$} & 20 {\scriptsize MHz},  30 {\scriptsize dBm}, -174 {\scriptsize dBm/Hz}, 3 {\scriptsize GHz}  \\\hline
        \multicolumn{1}{|l|} {$\sigma_{SF}, \quad h_{u,s}^{t}$} & $\mathcal{N}(0,\,4^{2}), \quad \mathcal{CN}(0,1)$\\\hline
		\multicolumn{1}{|l|} {$[v_{min},v_{max}], \tau_{max}$} & [1 m/s,4 m/s], 300 seconds \\\hline 
		\multicolumn{1}{|l|} {Slice data rate requirement, $R_{s}^{req}$} & eMBB: 10 Mbps;\newline URLLC: 250 kbps;\newline mMTC: 12 kbps \\\hline
        \multicolumn{1}{|l|} {$[f_{s}^{min},f_{s}^{max}]$ per slice} & eMBB: [0.005,0.5]; \newline URLLC: [0.0014,0.14]; \newline mMTC: [0.00047,0.047].  \\\hline
        \multicolumn{1}{|l|} {$[f^{min},f^{max}]$} & [0.01, 0.95] \\\hline
        \multicolumn{1}{|l|} {$\alpha, \quad \rho, \quad \zeta, \quad \Gamma_{th}$} & 0.5, 1.3, 5, 0.8  \\\hline
	\end{tabular}
 \end{center}
 \vspace{-3mm}
\end{table}

\vspace{-5mm}
\subsection{Training Performance}
\begin{figure}[t]
\captionsetup{aboveskip=0pt, belowskip=0pt}
\centering
	\includegraphics[width=0.9\columnwidth]{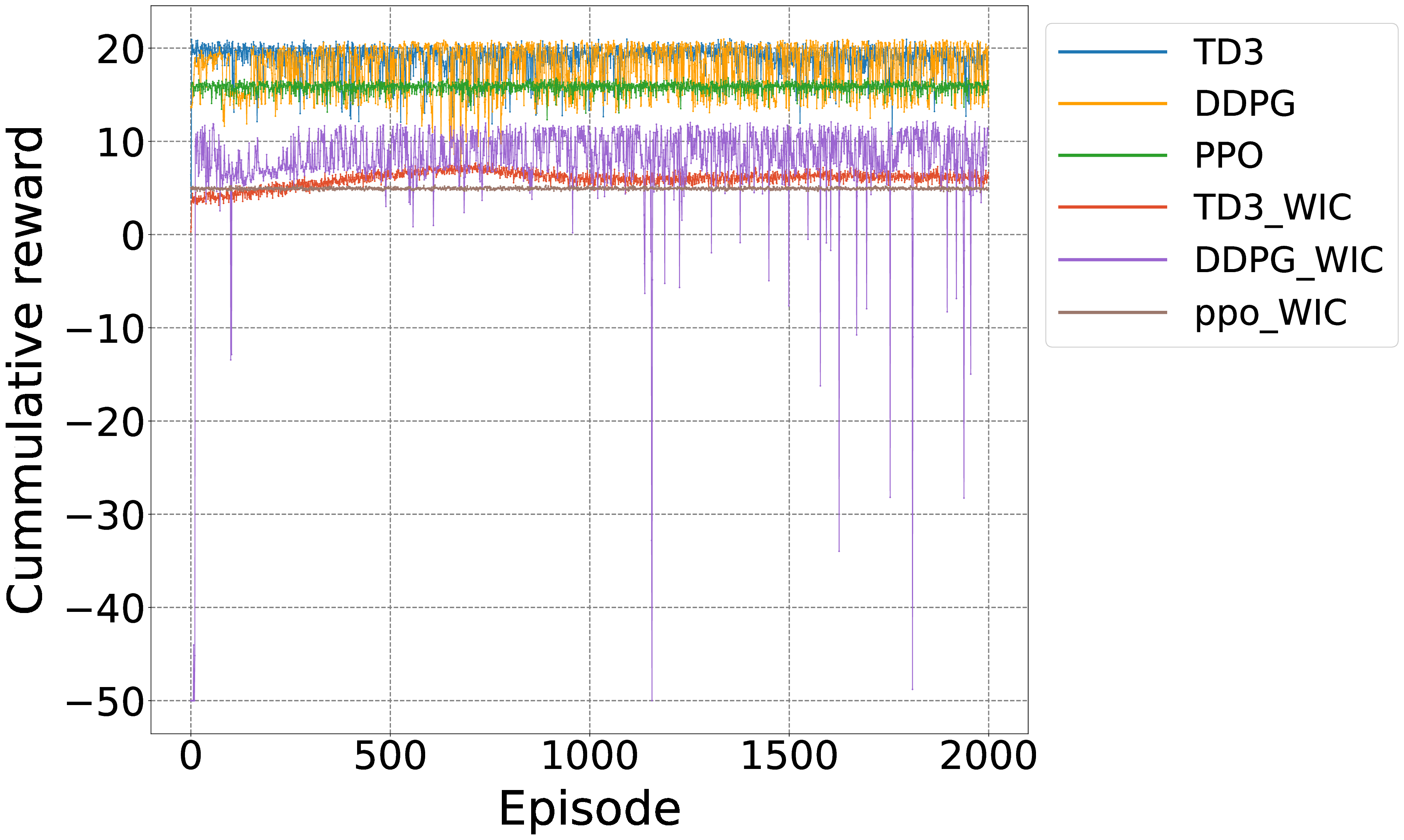}
	\caption{Inter-slice training performance.}
	\label{fig:03}
\end{figure}

\begin{figure*}[t]
    \centering
        \subfloat[eMBB agent.]{
            \includegraphics[width=0.64\columnwidth]{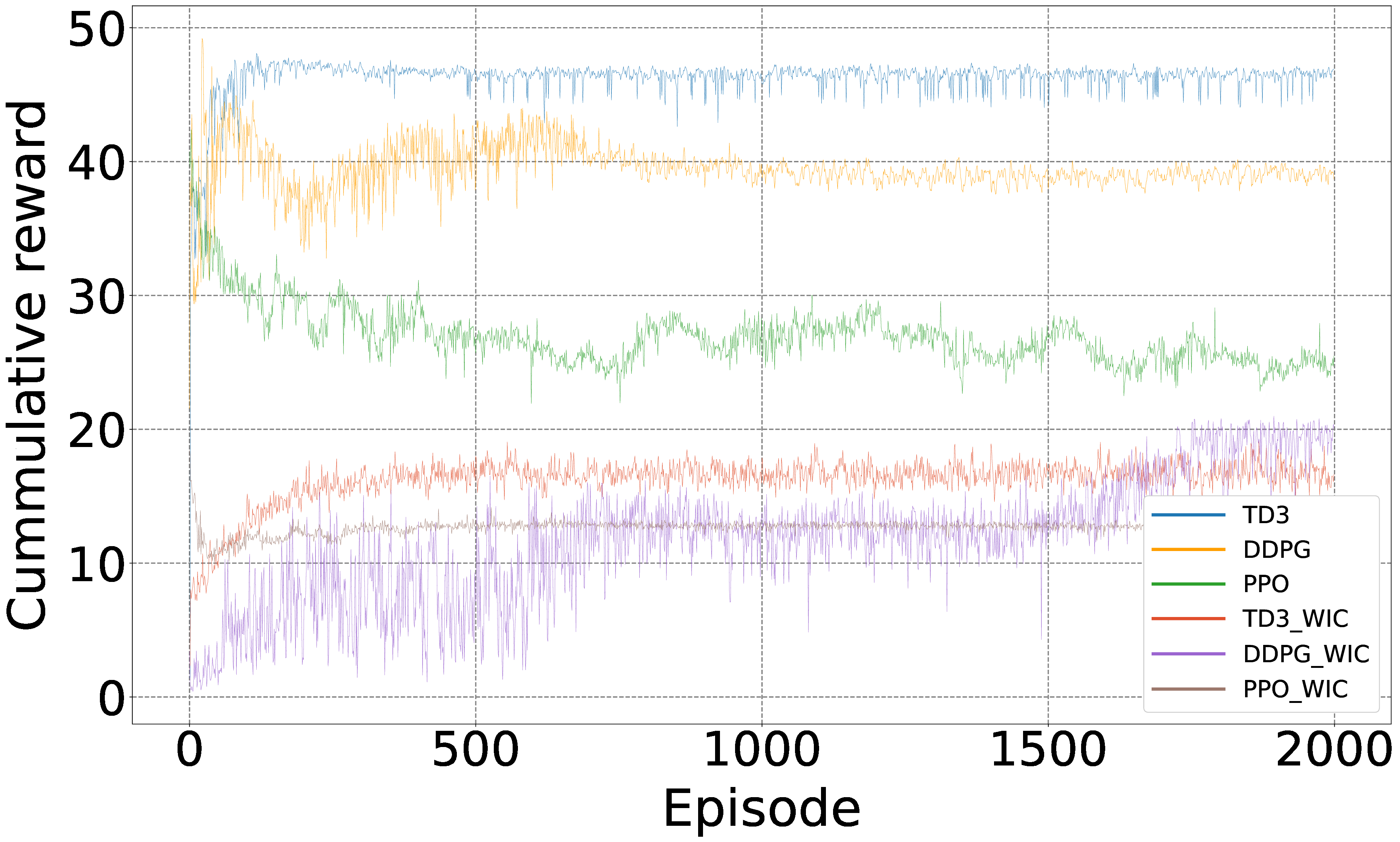}
            \label{fig:04_a}
        }
        \subfloat[URLLC agent.]{
            \includegraphics[width=0.64\columnwidth]{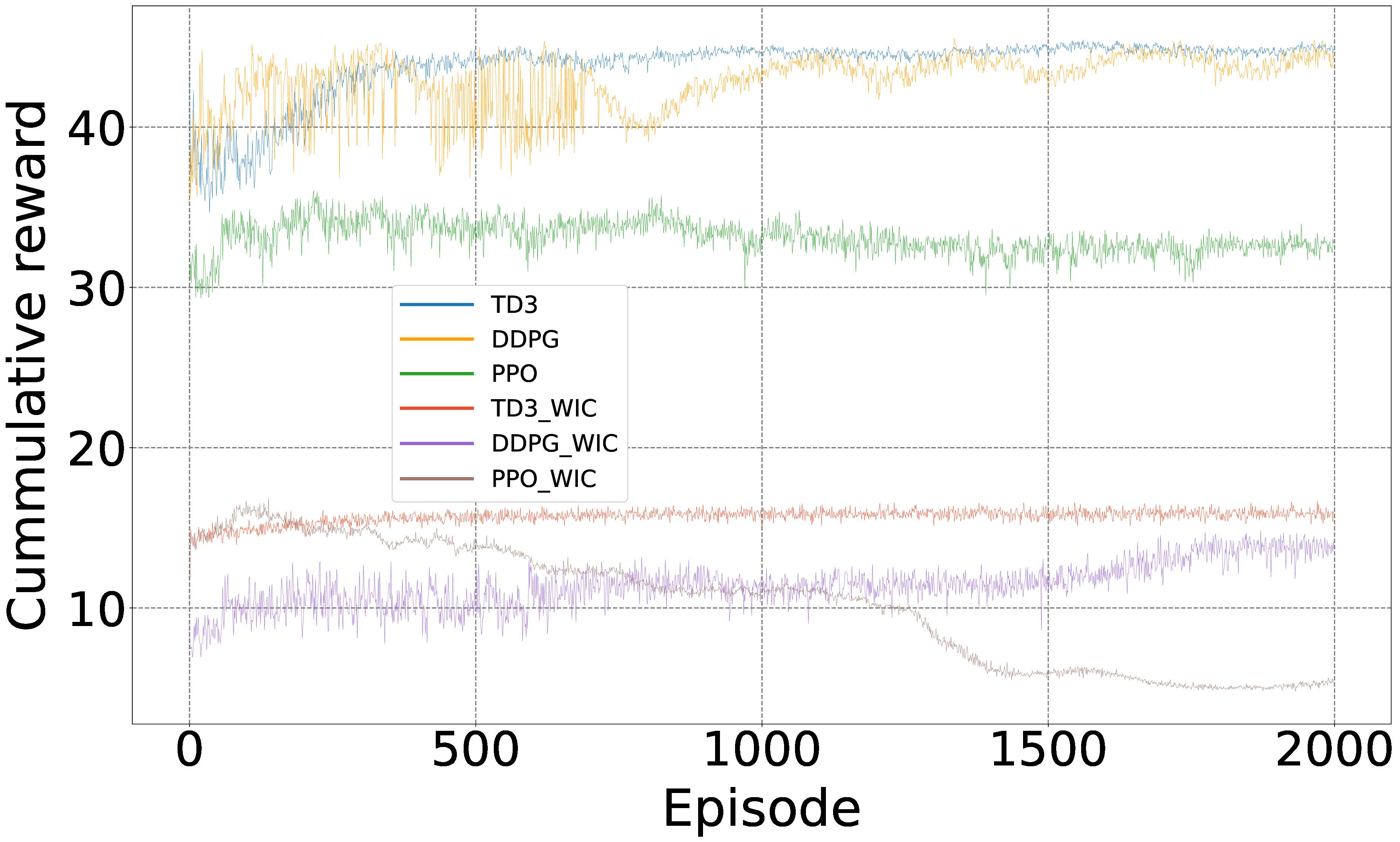}
            \label{fig:04_b}
        }
        \subfloat[mMTC agent.]{
            \includegraphics[width=0.64\columnwidth]{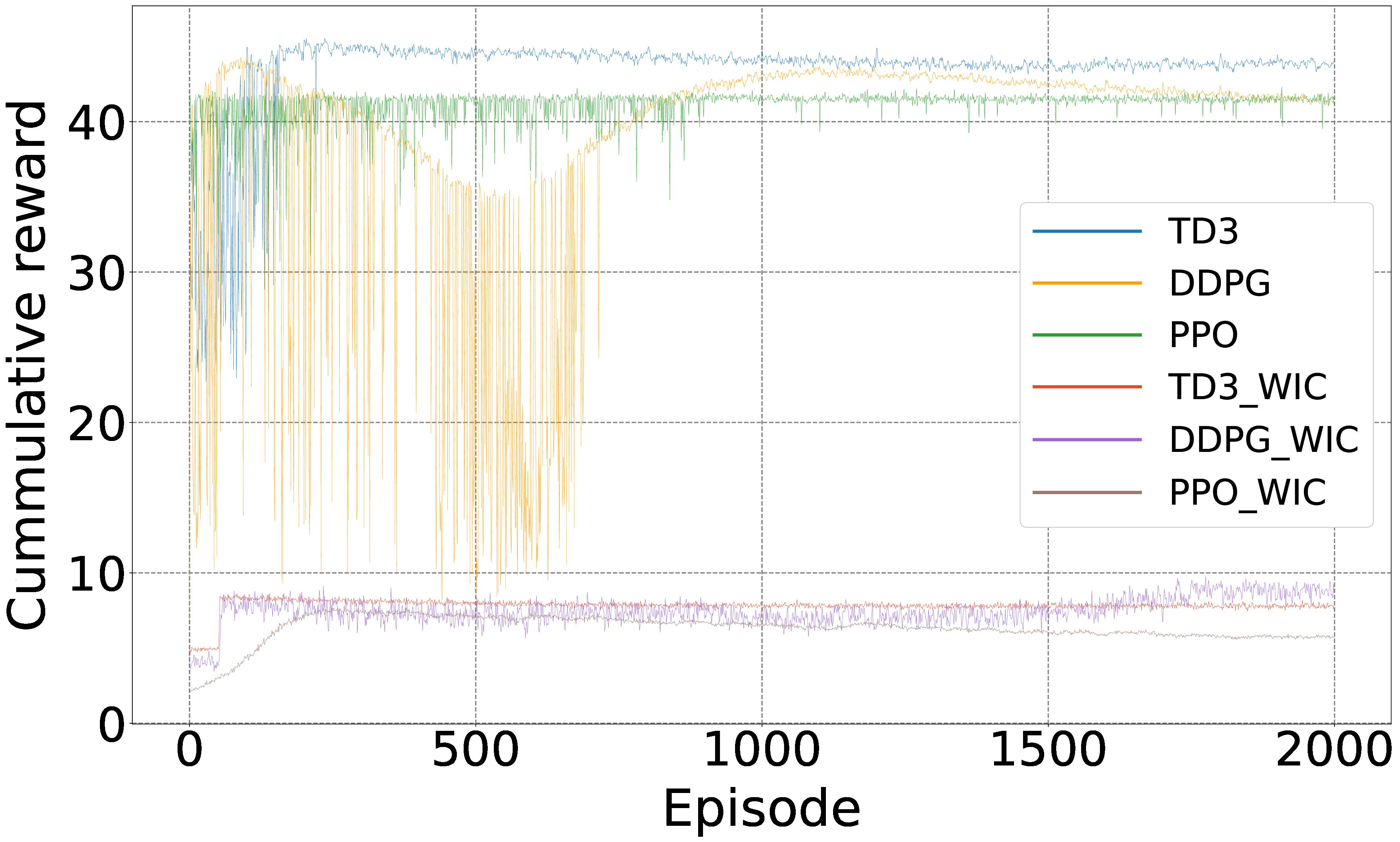}
            \label{fig:04_c}
        }
        \captionsetup{aboveskip=0pt, belowskip=0pt}
    \caption{Intra-slice training performance.}
    \label{fig:04}
    \vspace{-6mm}
\end{figure*}

\vspace{-1mm}
To evaluate the BS-IISI mechanism, we analyze the training performance of the DRL algorithms considered in this study and compare it with the performance of the selected baselines. Specifically, we examine, in Fig. \ref{fig:03}, \ref{fig:04_a}, \ref{fig:04_b}, and \ref{fig:04_c}, the reward behavior of each algorithm at each RAN slicing stage, \ie inter-slice bandwidth sharing and intra-slice bandwidth sharing. This analysis provides insights into the adaptability and optimization capabilities of the DRL algorithms in response to dynamic environmental conditions. The training performance results of the two RAN slicing stages mutually impact each other since the learning process of one is dependent on the bandwidth slicing results provided by the other. 

Fig. \ref{fig:03} illustrates the learning performance of the six DRL algorithms during the inter-slice bandwidth sharing stage over 2000 episodes of training. The results reveal that TD3 and DDPG perform very similarly, with a cumulative reward hovering around 20. However, DDPG exhibits more variance in its learning as compared to TD3. PPO also achieves a relatively stable performance, while the reward is slightly lower than that of TD3 and DDPG. This can be explained by the fact that PPO relies on recent data, which limits its ability to learn from past experience, potentially diminishing its performance compared to off-policy algorithms.
 TD3-WIC, DDPG-WIC, and PPO-WIC present a more volatile behavior. In particular, the cumulative reward of DDPG-WIC significantly fluctuates with occasional negative spikes. Without isolation constraints, TD3-WIC, DDPG-WIC, and PPO-WIC explore a wider range of actions, which leads to unstable rewards and, consequently, less efficient bandwidth allocation to slices.
\begin{table}[t]
    \centering
    \caption{Training parameters.}
    \label{tab:3}
    \begin{tabular}{|m{2.3cm}|m{1.7cm}|m{1.7cm}|m{1.4cm}|}
        \hline
        \textbf{Hyperparameter} & \textbf{TD3} & \textbf{DDPG} & \textbf{PPO} \\ \hline
        {\scriptsize No. of hidden layers} & 2 & 2 & 2 \\ \hline
        {\scriptsize No. of hidden units/layer} & \parbox{2cm}{\tiny{Global agent: 300/200} \\ \tiny{Slice agent: 500/400}} & \parbox{2cm}{\tiny{Global agent: 300/200} \\ \tiny{Slice agent: 500/400}} & \parbox{1.6cm}{\tiny{Global agent: 300/200} \\ \tiny{Slice agent: 500/400}} \\ \hline
        {\scriptsize Actor learning rate} & 0.0001 & 0.0001 & 0.0001 \\ \hline
        {\scriptsize Critic learning rate} & 0.001 & 0.001 & 0.001 \\ \hline
        {\scriptsize Replay buffer size} & \parbox{2cm}{\tiny{Global agent: $10^{5}$ \\ Slice agent: $10^{6}$}} & \parbox{2cm}{\tiny{Global agent: $10^{5}$ \\ Slice agent: $10^{6}$}} & None \\ \hline
        {\scriptsize Batch size} & 128 & 128 & 500 \\ \hline
        {\scriptsize Optimizer} & Adam & Adam & Adam \\ \hline
        {\scriptsize Activation function} & ReLU & ReLU & ReLU \\ \hline
        {\scriptsize Exploration noise distribution} & \parbox{2.5cm}{\tiny{Global agent:} $\mathcal{N}(0,0.2)$ \\ \tiny{eMBB agent:} $\mathcal{N}(0,0.1)$ \\ \tiny{URLLC agent:} $\mathcal{N}(0,0.1)$ \\ \tiny{mMTC agent:} $\mathcal{N}(0,0.1)$} & \parbox{2.5cm}{\tiny{Global agent:} $OU(0,0.2)$ \\\tiny{eMBB agent:} $OU(0,0.5)$ \\ \tiny{URLLC agent:} $OU(0,0.1)$ \\ \tiny{mMTC agent:} $OU(0,0.1)$} & None \\ \hline
        {\scriptsize No. of epochs} & None & None & 10 \\ \hline
        {\scriptsize Target net. update rate} & 0.001 & 0.001 & None \\ \hline
        {\scriptsize Target net. update freq.} & 10 & 10 & None \\ \hline
        {\scriptsize Discount factor} & 0.99 & 0.99 & 0.99 \\ \hline
        {\scriptsize No. of steps/episode} & 50 & 50 & 50 \\ \hline
        {\scriptsize Clipping parameter} & None & None & 0.2 \\ \hline
    \end{tabular}
\end{table}


Fig. \ref{fig:04_a}, \ref{fig:04_b} and \ref{fig:04_c} depict the reward values of the six DRL algorithms for the eMBB slice agent, the URLLC slice agent, and the mMTC slice agent, respectively. The TD3 algorithm performs well across all slices, consistently achieving high cumulative rewards with less fluctuation. The DDPG algorithm shows a lower performance as compared to TD3, especially for the eMBB slice where the user requirements in terms of data rate are very high. In addition, DDPG exhibits high instability, with large fluctuations during the early training episodes before stabilizing, especially for the mMTC slice. This outcome is due to the high variance of DDPG algorithm policy updates. However, the TD3 algorithm uses a more stable actor-critic approach with delayed policy updates and target networks. PPO demonstrates a relatively high reward during early training episodes, which is however followed by a decline to stabilize at a lower cumulative reward than TD3 and DDPG. This can occur because the PPO algorithm samples actions from a stochastic policy, which frequently leads to violations of intra-slice isolation constraints. TD3-WIC, DDPG-WIC, and PPO-WIC perform worse than their respective constrained algorithms across all slices. They show a slower progression during the early training episodes, but quickly stabilize at a very low cumulative reward, especially the PPO-WIC algorithm. This explains that, in the absence of isolation constraints, learning efficiency can be compromised.

\vspace{-8mm}
\subsection{Inter-Slice Bandwidth Sharing Performance}
\vspace{-1mm}
To evaluate inter-slice bandwidth allocation, we compare the performance of all algorithms using the objective function and the reconfiguration cost defined in Eq. \eqref{eq:obj1}  and \eqref{eq:cost_total}, respectively. For the RSSI-IP mechanism, the objective function value excludes the reconfiguration cost, as it is not considered in \cite{yarkina2022multi}. Each algorithm is evaluated under various network configurations, particularly varying the number of users. The DRL algorithm's inference model was trained on a fixed number of users per slice: 20 for eMBB, 70 for URLLC, and 210 for mMTC. This enables an evaluation of the algorithms' robustness in maintaining effective performance despite changes in environmental conditions.

\begin{figure}[t]
    \centering
	\includegraphics[width=0.9\columnwidth]{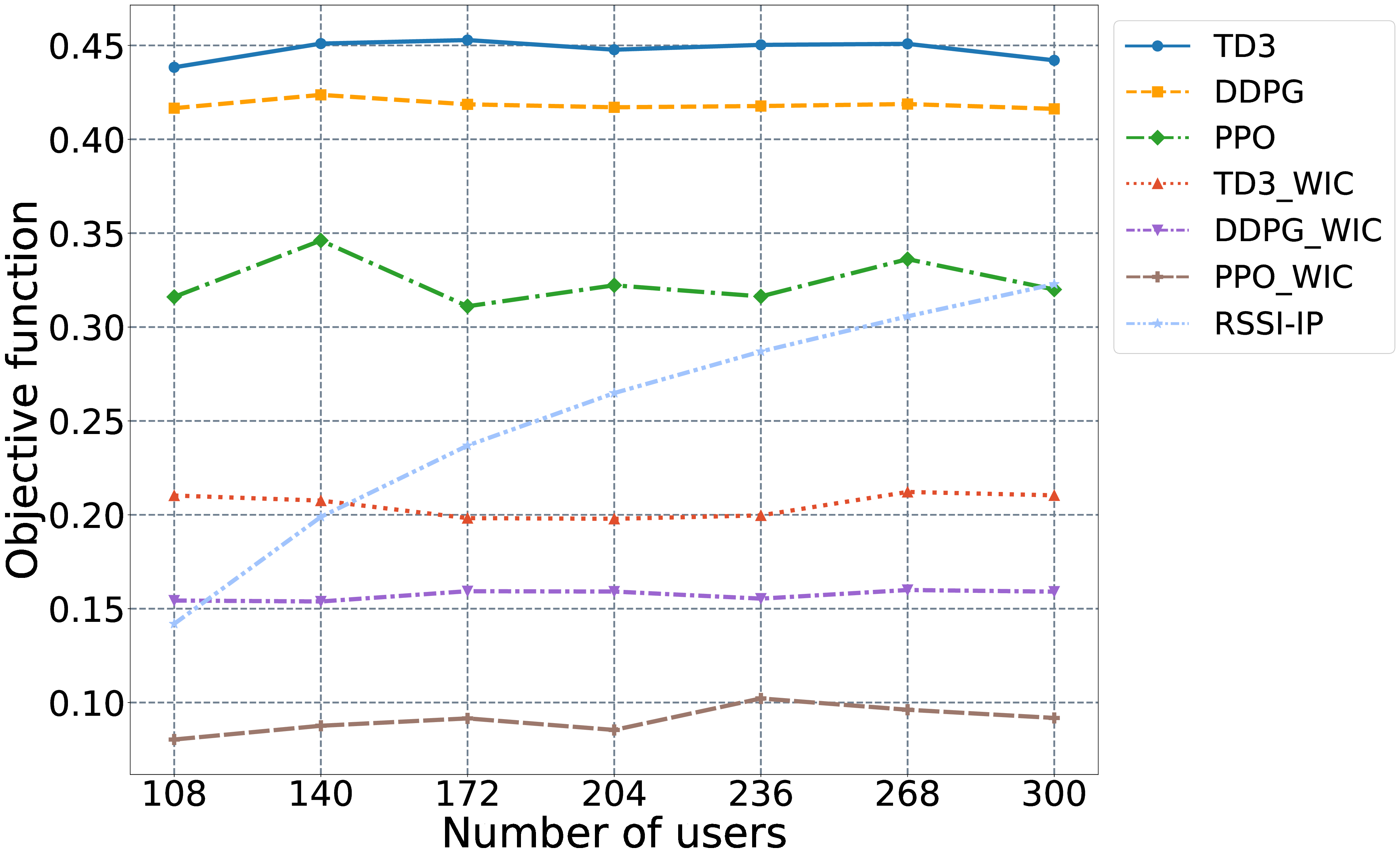}
	\caption{Objective function Eq. \eqref{eq:obj1}}
	\label{fig:05}
    \vspace{-2mm}
\end{figure}

Fig. \ref{fig:05} illustrates the behavior of the objective function under varying system configurations, represented by different numbers of users. Among the evaluated algorithms, TD3 algorithm consistently achieves the highest objective function values across all configurations, confirming its effectiveness in optimizing the objective function. Furthermore, its stable performance as the number of users changes demonstrates its robustness to scaling. DDPG performs slightly below the TD3, but still remains consistent across all configurations. In its turn, the PPO algorithm shows moderate performance, with some fluctuation in its behavior when the number of users varies, indicating less stability as compared to TD3 and DDPG. 

\begin{figure}[t]
    \centering
	\includegraphics[width=0.9\columnwidth]{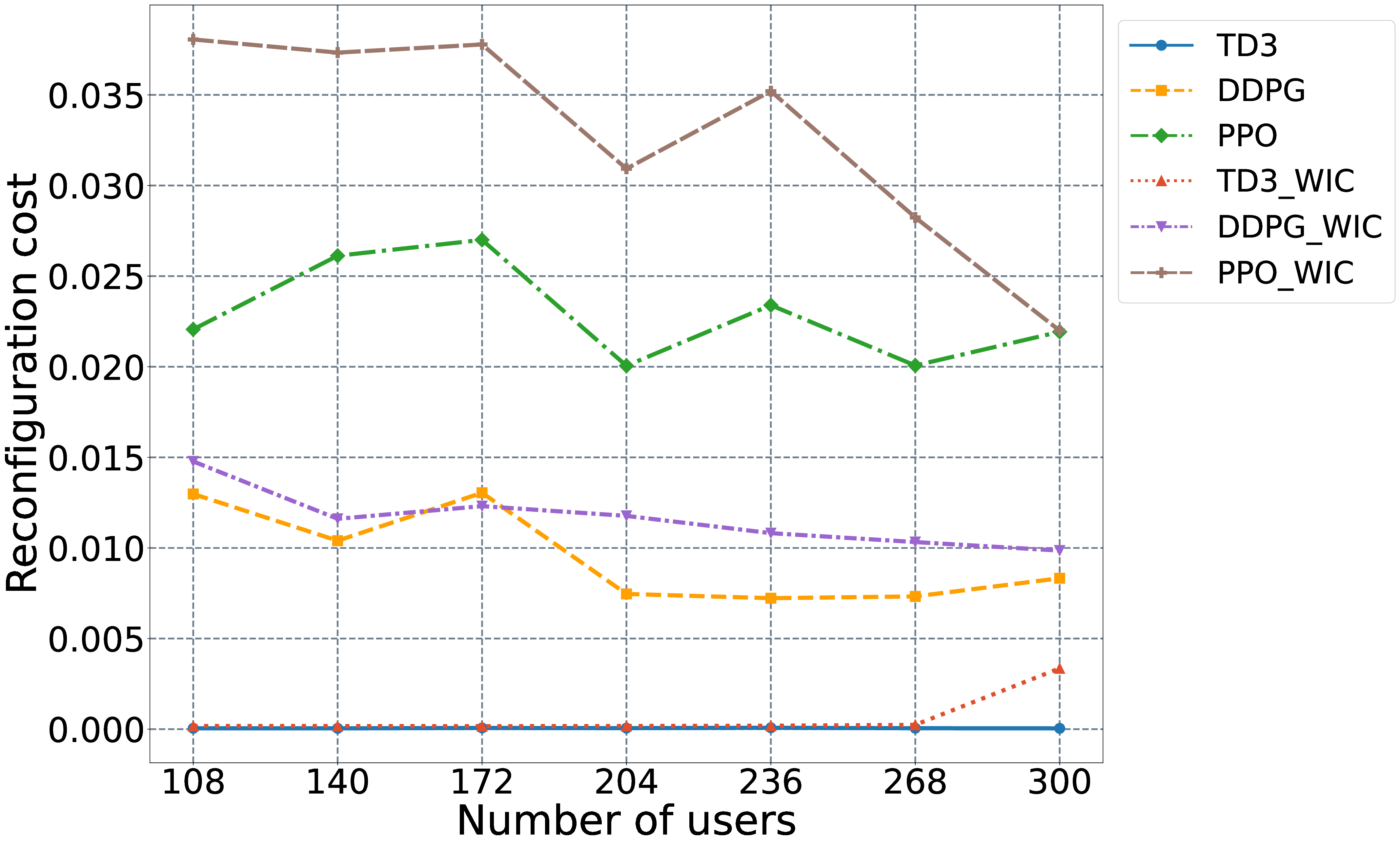}
	\caption{Total reconfiguration cost Eq. \eqref{eq:cost_total}.}
	\label{fig:06}
    \vspace{-2.5mm}
\end{figure}

RSSI-IP outperforms the TD3-WIC, DDPG-WIC, and PPO-WIC under most network configurations, but falls short as compared to TD3, DDPG, and PPO. RSSI-IP prioritizes generous resource allocation to meet users' minimum required data rates, disregarding potential resource wastage. Its performance steadily increases with the number of users: specifically, when users are fewer, abundant resources cause over-allocation, reducing satisfaction due to penalization for resource wastage in the utility function defined in Eq. \eqref{eq:sd_user}. This explains why the system's total degree of satisfaction is significantly lower for 108 users than for 300 users. The TD3-WIC, DDPG-WIC, and PPO-WIC algorithms consistently underperform, which emphasizes the importance of isolation constraints.

Fig. \ref{fig:06} shows the results of comparing the reconfiguration cost defined in Eq. \eqref{eq:cost_total} generated by the TD3, DDPG, PPO, TD3-WIC, DDPG-WIC, and PPO-WIC algorithms for different numbers of users. The results reveal that the isolation constraints considered in TD3, DDPG, and PPO lead to reduced reconfiguration costs as compared to their respective unconstrained algorithms, i.e., TD3-WIC, DDPG-WIC, and PPO-WIC. This demonstrates the importance of considering isolation constraints to enhance inter-slice isolation. The TD3, DDPG, and PPO algorithms maintain a relatively stable performance regardless of the number of users, which highlights their robustness in handling scenarios of different sizes.


\vspace{-2mm}
\subsection{Intra-Slice Bandwidth Sharing Performance}
To evaluate intra-slice bandwidth allocation, we analyze the performance of the selected algorithms for the eMBB, URLLC, and mMTC slices in terms of users' achieved data rate, degree of satisfaction, and resource wastage in the bandwidth allocation process. This analysis provides important insights into how effectively each algorithm allocates bandwidth among different groups of users, each with distinct requirements. In addition, the resource wastage metric evaluates the efficiency of bandwidth sharing across users.

\begin{figure}[t]
     \centering
     \begin{subfigure}[b]{0.7\columnwidth}
         \centering
         \captionsetup{aboveskip=0pt, belowskip=0pt}\includegraphics[width=\columnwidth]{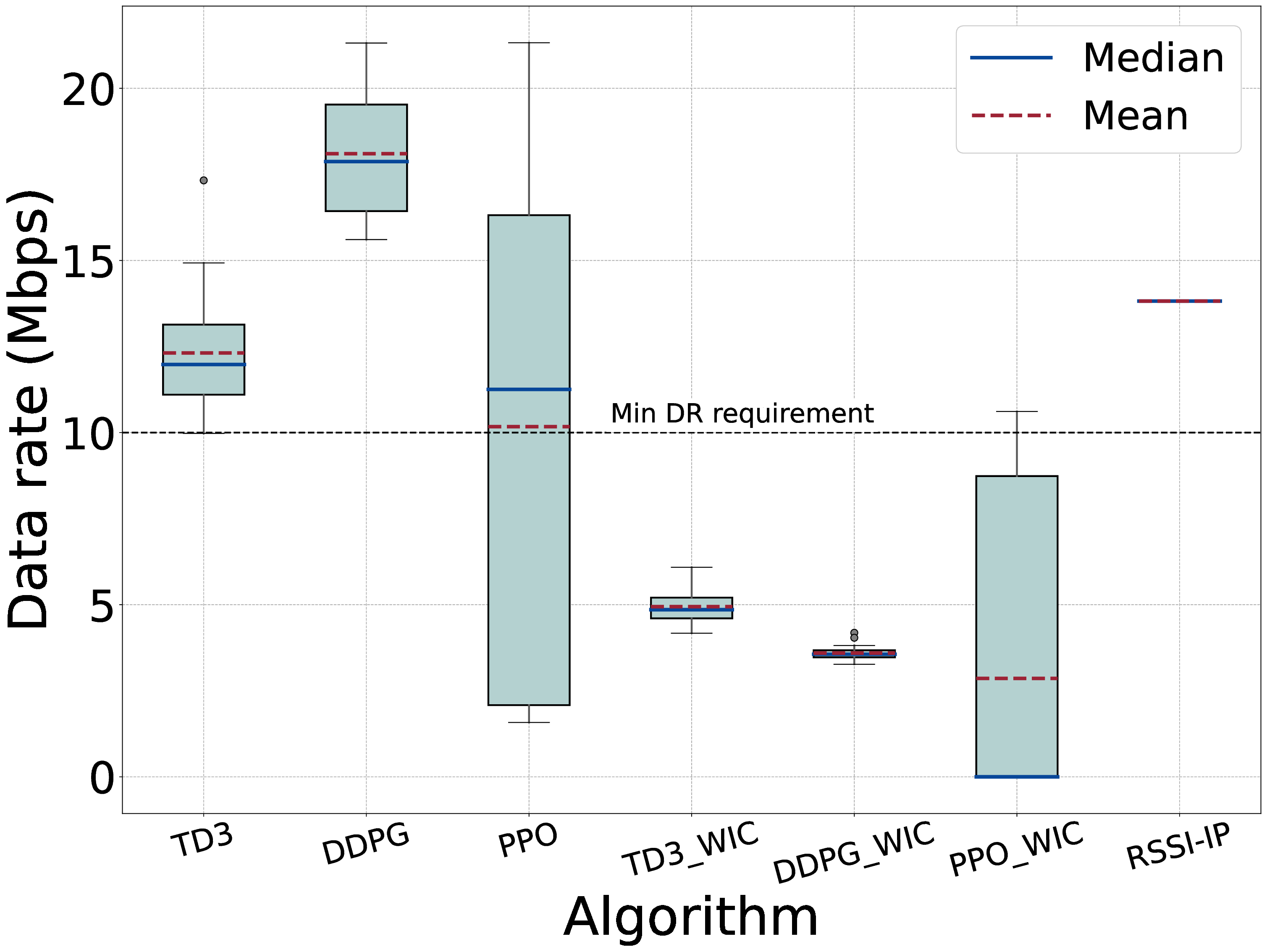}
         \caption{Data rate of users Eq. \eqref{eq:datarate}.}
         \label{fig:07_a}
     \end{subfigure}
     \begin{subfigure}[b]{0.7\columnwidth}
         \centering
         \captionsetup{aboveskip=0pt, belowskip=0pt}\includegraphics[width=\columnwidth]{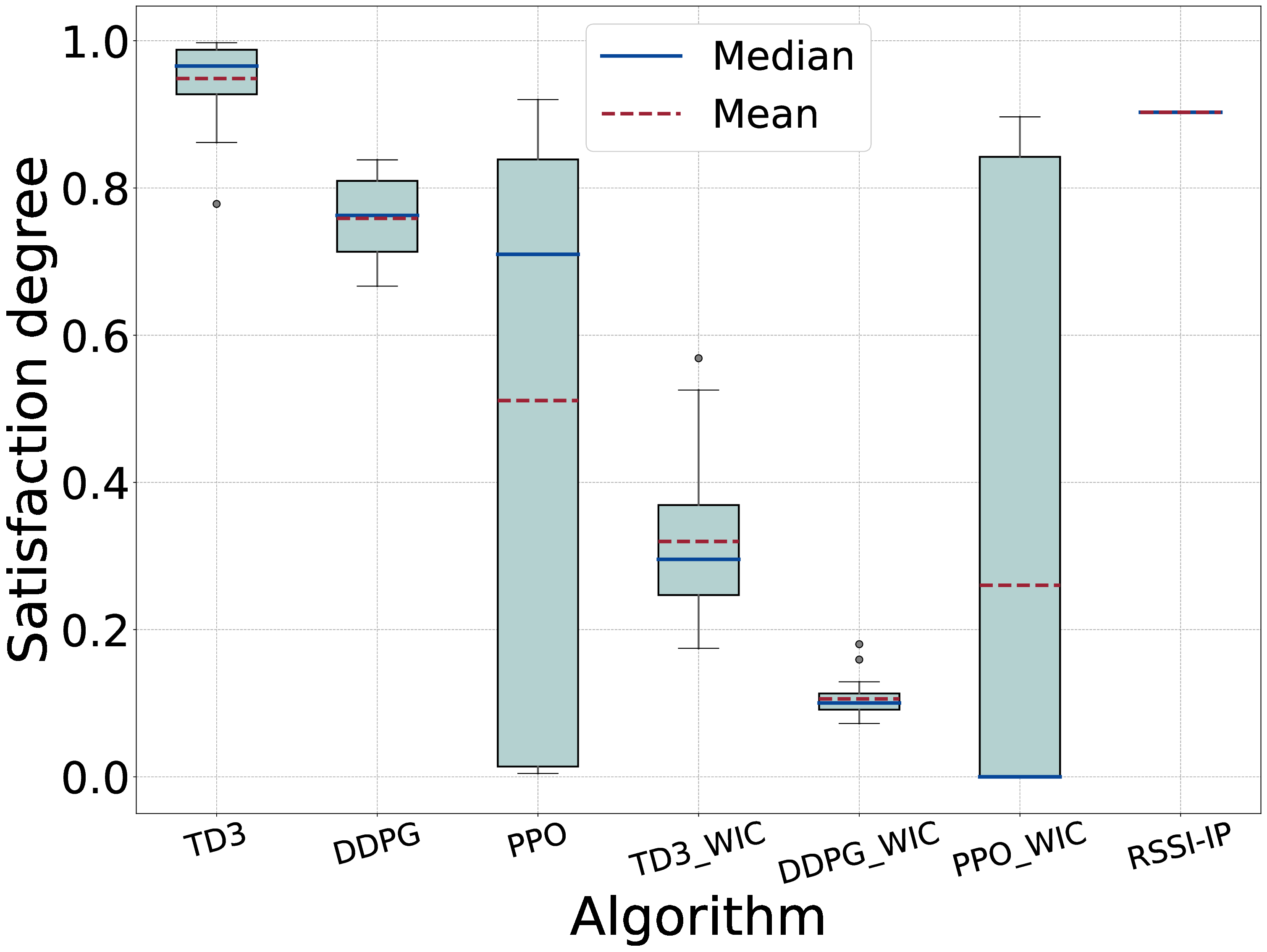}
         \caption{Satisfaction degree of users Eq. \eqref{eq:sd_user}.}
         \label{fig:07_b}
     \end{subfigure}
        \captionsetup{aboveskip=0pt, belowskip=0pt}
        \caption{QoS achieved by eMBB users.}
        \label{fig:07}
        \vspace{-1.5mm}
\end{figure}

\begin{figure}[h]
     \centering
     \begin{subfigure}[b]{0.7\columnwidth}
         \centering
         \captionsetup{aboveskip=0pt, belowskip=0pt}\includegraphics[width=\columnwidth]{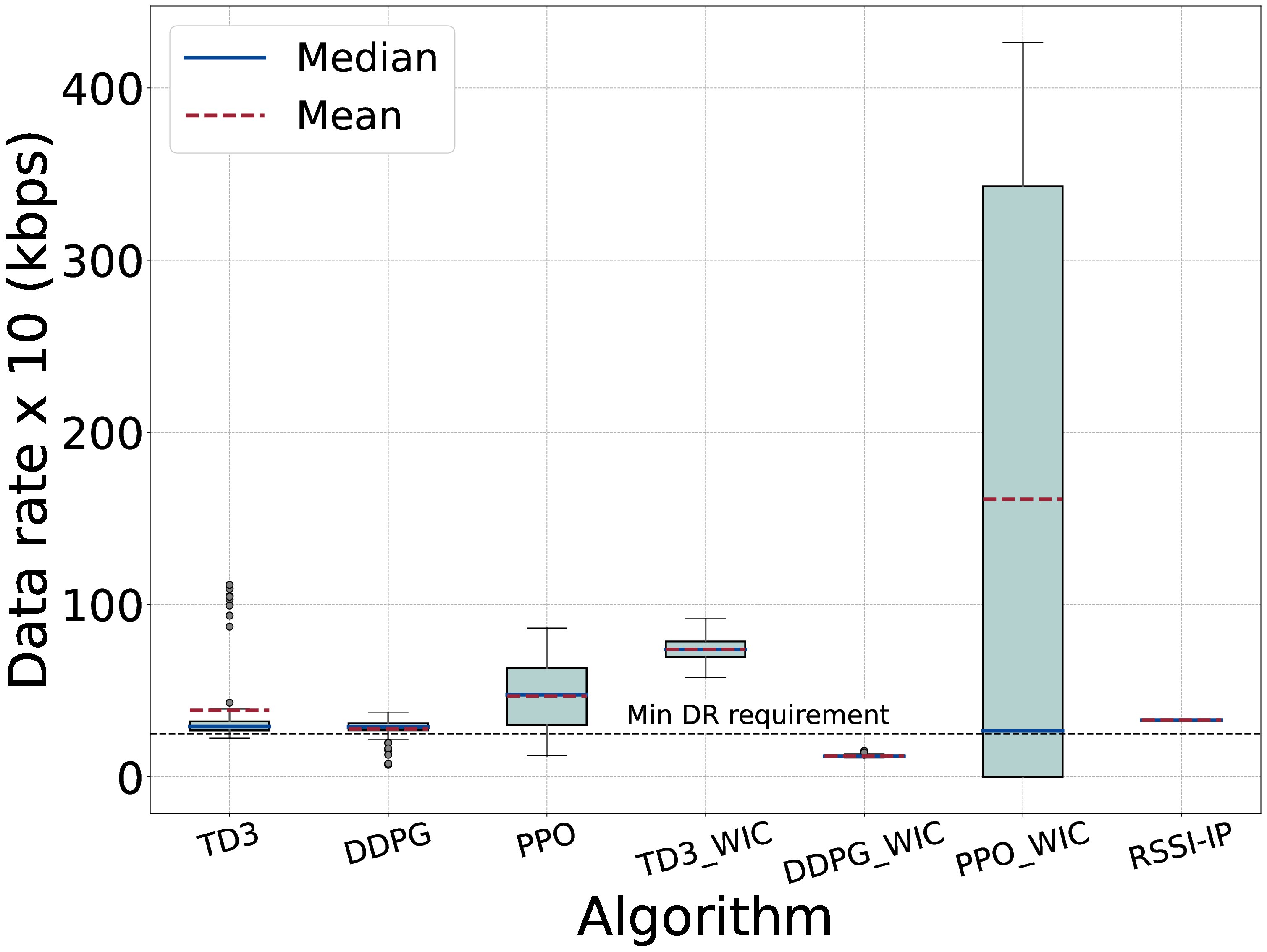}
         \caption{Data rate of users Eq. \eqref{eq:datarate}.}
         \label{fig:08_a}
     \end{subfigure}
     \begin{subfigure}[b]{0.7\columnwidth}
         \centering
         \captionsetup{aboveskip=0pt, belowskip=0pt}\includegraphics[width=\columnwidth]{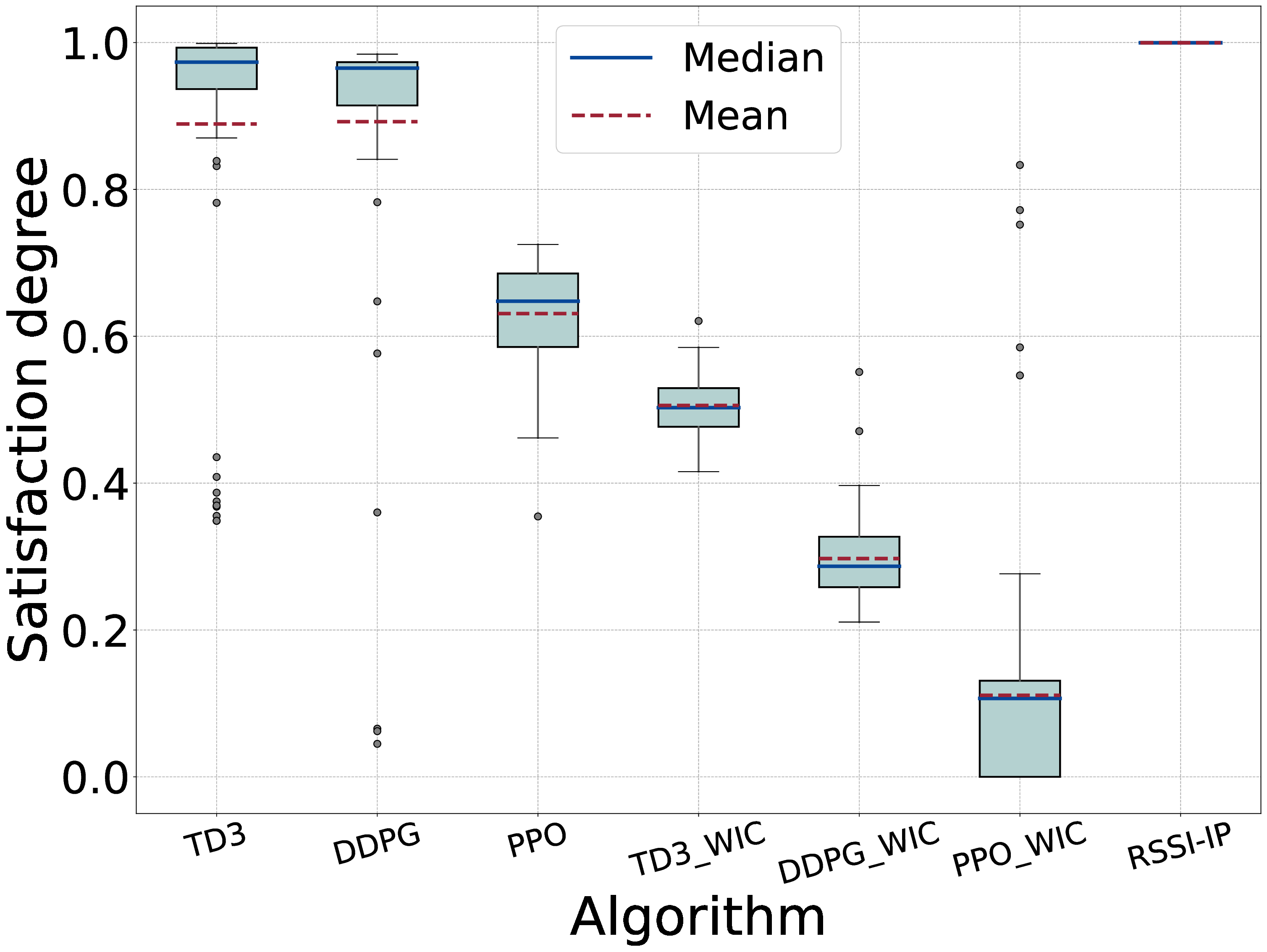}
         \caption{Satisfaction degree of users Eq. \eqref{eq:sd_user}.}
         \label{fig:08_b}
     \end{subfigure}
        \captionsetup{aboveskip=0pt, belowskip=0pt}
        \caption{QoS achieved by URLLC users.}
        \label{fig:08}
        \vspace{-1.5mm}
\end{figure}

\begin{figure}[t]
     \centering
     \begin{subfigure}[b]{0.7\columnwidth}
         \centering
         \captionsetup{aboveskip=0pt, belowskip=0pt}\includegraphics[width=\columnwidth]{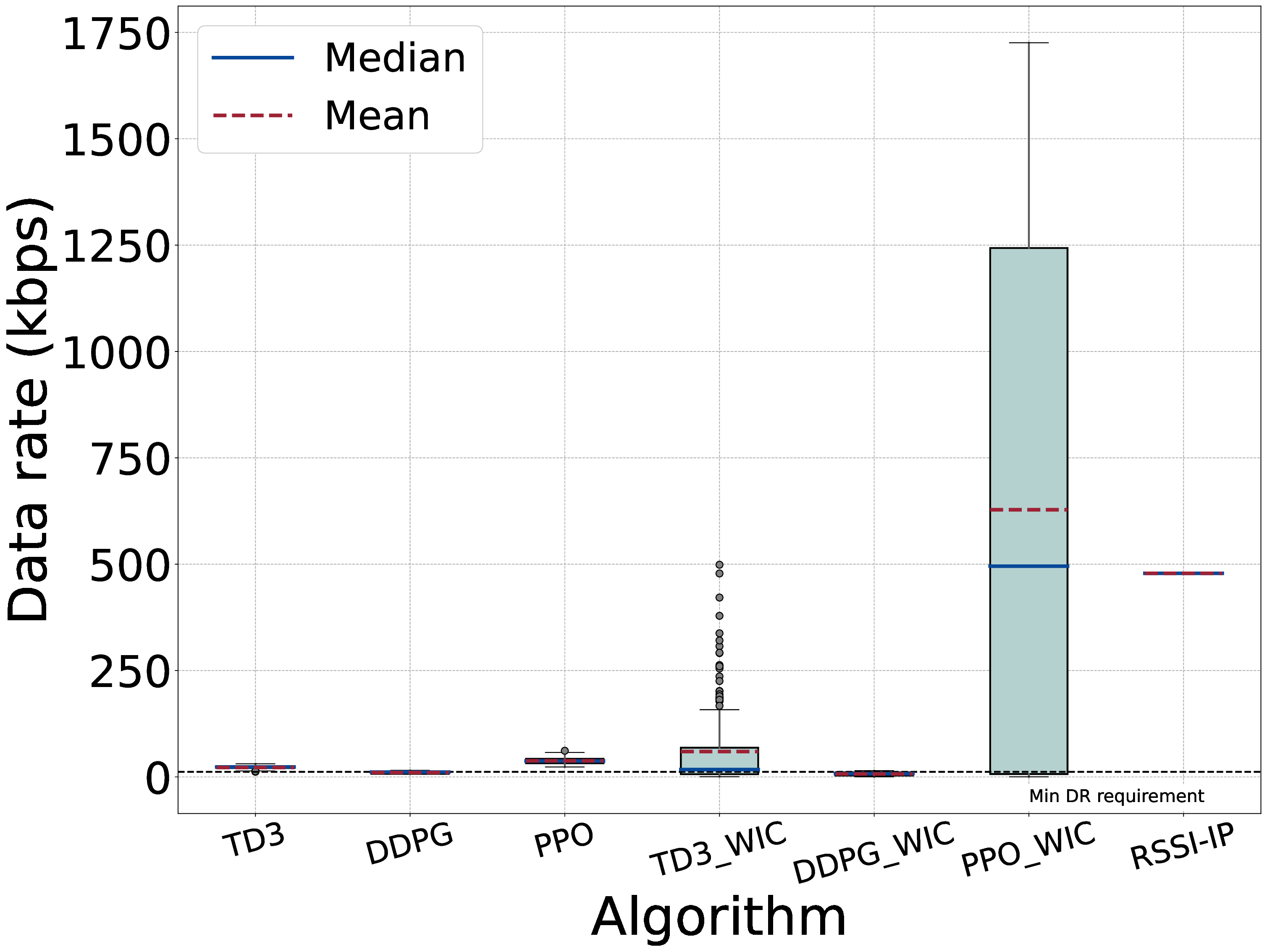}
         \caption{Data rate of users Eq. \eqref{eq:datarate}.}
         \label{fig:09_a}
     \end{subfigure}
     \begin{subfigure}[b]{0.7\columnwidth}
         \centering
         \captionsetup{aboveskip=0pt, belowskip=0pt}\includegraphics[width=\columnwidth]{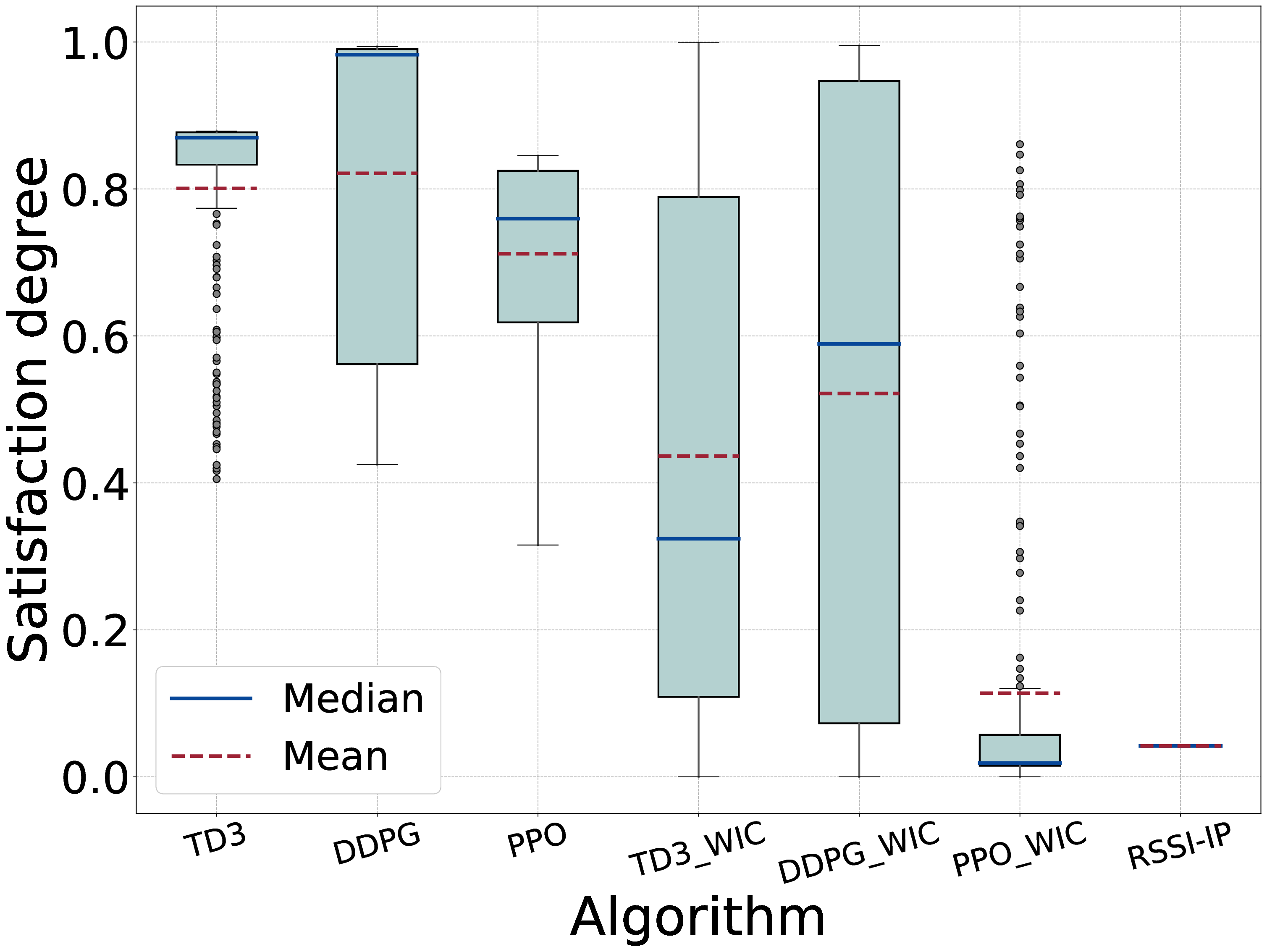}
         \caption{Satisfaction degree of users Eq. \eqref{eq:sd_user}.}
         \label{fig:09_b}
     \end{subfigure}
        \captionsetup{aboveskip=0pt, belowskip=0pt}
        \caption{QoS achieved by mMTC users.}
        \label{fig:09}
        \vspace{-1.5mm}
\end{figure}

\vspace{0pt}
Figs. \ref{fig:07}, \ref{fig:08} and \ref{fig:09} show the box plots representing the distribution of the data rate and degree of satisfaction of 20 eMBB users, 70 URLLC users and 210 mMTC users, respectively.
We arrive at the following observations from the results presented in those figures.
    For the eMBB slice, the TD3 algorithm achieves the highest median degree of satisfaction, close to 1, with a small interquartile range (IQR) (see Fig. \ref{fig:07_b}). This reflects its ability to efficiently allocate bandwidth, ensuring that users' data rates are consistently near the minimum required threshold (10 Mbps). The RSSI-IP algorithm’s median degree of satisfaction is slightly lower than that of TD3. Its box plot is flat, as resources are uniformly allocated among the slice’s users. The DDPG algorithm performs slightly worse than the TD3 and RSSI-IP algorithms. This is due to the excessive bandwidth allocated to users, i.e., DDPG has the highest median data rate (see Fig. \ref{fig:07_a}), which is penalized by the utility function defined in Eq. \eqref{eq:sd_user}. DDPG has a larger IQR than TD3 and RSSI-IP, indicating more variability in satisfaction across users. However, this variability is still controlled and has no significant impact on the overall performance. Furthermore, the PPO algorithm exhibits a wide dispersion of users' data rates, leading to frequent over- or under-allocation of bandwidth and a corresponding drop in users' satisfaction. For the TD3-WIC, DDPG-WIC, and PPO-WIC algorithms, the absence of isolation constraints results in a significant drop in the users' satisfaction.
    
    \vspace{0pt}
    For the URLLC slice, the TD3 algorithm achieves significantly higher data rates for some users as compared to the RSSI-IP (see Fig. \ref{fig:08_a}), resulting in a higher mean data rate above the minimum threshold. Consequently, RSSI-IP outperforms TD3 in user satisfaction. Despite this, TD3 guarantees near-perfect user satisfaction, with an IQR between 0.9 and 1 (see Fig. \ref{fig:08_b}), demonstrating consistent satisfaction for most users. DDPG has several outliers below the required minimum data rate. While its median and mean data rates might appear comparable to or better than TD3, these outliers reveal potential issues for a subset of users, which makes DDPG less fair and consistent than TD3. The PPO algorithm’s data rate box plot shows a larger IQR above the required minimum threshold compared to RSSI-IP, TD3, and DDPG, indicating a greater dispersion in the data rates provided to URLLC users. This variability leads to reduced user satisfaction. The TD3-WIC, DDPG-WIC, and PPO-WIC algorithms perform poorly overall, with frequent instances of data rates falling significantly below the required threshold (e.g., DDPG-WIC) or exhibiting a very large spread in data rates (e.g., PPO-WIC).

    For the mMTC slice, the DDPG algorithm performs better than TD3, achieving a median degree of satisfaction close to 1 (see Fig. \ref{fig:09_b}). However, its IQR is larger than that of TD3, indicating significant variability in user satisfaction. By contrast, TD3's satisfaction distribution includes lower quartiles and outliers, reflecting more fluctuating performance for some users. The PPO algorithm achieves a median satisfaction of approximately 0.7, which is lower than both DDPG and TD3. While its degree of satisfaction IQR is narrower than that of DDPG, the lower median suggests that most users experience less satisfaction as compared to the DDPG and TD3 algorithms. The RSSI-IP algorithm allocates resources generously, as shown in its data rate plot (see Fig. \ref{fig:09_a}). However, this over-allocation of resources results in a low degree of satisfaction for users. DDPG-WIC reaches comparable data rates to DDPG and TD3, but most values remain below the required threshold for users. Consequently, DDPG-WIC has a clearly lower median satisfaction ($\sim$0.6) as compared to DDPG ($\sim$1.0) and TD3 ($\sim$0.85). The PPO-WIC algorithm experiences significant difficulties with resource allocation, which results in a large over-allocation of bandwidth. Therefore, PPO-WIC has the lowest user satisfaction among all the algorithms, with a median satisfaction of approximately 0.05.

\begin{figure}[t]
\centering
	\captionsetup{aboveskip=0pt, belowskip=0pt}\includegraphics[width=0.9\columnwidth]{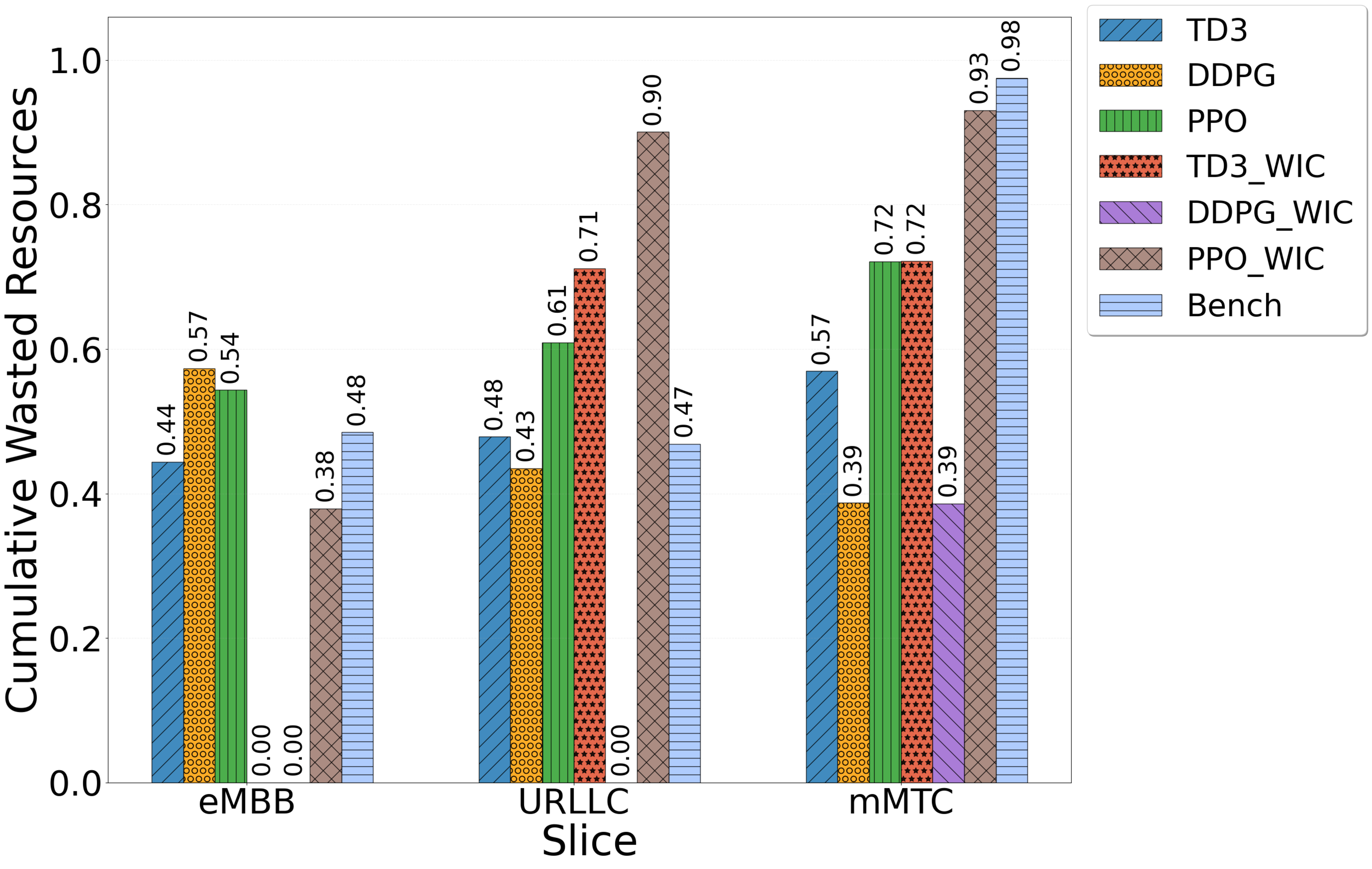}
	\caption{Resource wastage by slice.}
	\label{fig:10}
    \vspace{-1.5mm}
\end{figure}

Fig. \ref{fig:10} plots the average rate of resource wastage in each RAN slice. We define resource wastage for a user based on how much its achieved data rate exceeds the minimum required data rate. For a user $u \in \mathscr{U}_s, \forall s \in \mathscr{S}$, if $r_{u,s}^{t} > R_{s}^{req}$, we calculate resource wastage as follows: $wst_{u,s}^{t} = e^{-\frac{R_{s}^{req}}{r_{u,s}^{t}}}$. The average rate of resource wastage in slice $s$ is given by: $wst_{s}^{t}=\frac{1}{|\mathscr{U}_s|}\sum_{u \in \mathscr{U}_s} wst_{u,s}^{t}$. 

\vspace{1mm}
For the eMBB slice, we observe that the TD3-WIC and DDPG-WIC algorithms did not waste any resources. This is because none of the slice's users reached the minimum data rate required. The TD3 algorithm, while showing some resource wastage, achieves the highest user’s degree of satisfaction (see Fig. \ref{fig:07_b}), balancing resource efficiency and user satisfaction. The resource wastage results for the URLLC and mMTC slices reflect the results obtained in Fig. \ref{fig:08_a} and \ref{fig:09_a}, respectively. Although other algorithms like PPO, DDPG, and RSSI-IP show variability in resource wastage, TD3 maintains a relatively stable resource wastage rate across all slices, highlighting its robustness and adaptability.

\vspace{-2mm}
\section{Conclusion and Future Work}
\vspace{-1.5mm}

This paper introduces the BS-IISI mechanism, a RAN slicing solution for TNs designed to balance dynamic bandwidth sharing with robust inter- and intra-slice isolation. The mechanism operates in two stages: inter-slice and intra-slice bandwidth sharing. In the first stage, each slice receives a portion of the system’s total bandwidth. In the second stage, each slice partitions its allocated bandwidth among its users. The bandwidth allocation problem is formulated as an optimization task and solved using DRL algorithms. The BS-IISI mechanism uses DRL in each stage to dynamically allocate bandwidth while maintaining isolation through resource reconfiguration constraints related to QoS performance. We proposed to deploy the BS-IISI mechanism within an O-RAN architecture, detailing the functional blocks supporting its operation and the lifecycle management of the DRL models used. To evaluate the performance of the BS-IISI mechanism, we implemented the following three variations using different DRL algorithms: TD3, DDPG, and PPO. These implementations were then compared against several baselines, showing that the TD3 algorithm outperforms both the baselines and the other DRL algorithms.

In future work, we will consider that a user can be associated with multiple slices and incorporate diverse user mobility models. In addition, we will investigate techniques to accurately monitor DRL inference models used in RAN slicing and detect when they suffer from a severe degradation in performance so they can be proactively replaced by appropriate new models.

\vspace{-6mm}





\bibliographystyle{IEEEtran}
\bibliography{IEEEabrv,ref}

\end{document}